\documentclass[preprint]{emulateapj}

\newcommand{\fer}{{\it Fermi}}
\newcommand{\wse}{{\it WISE}}
\newcommand{\strip}{{\it WISE blazar Strip}}

\usepackage{graphicx}
\usepackage{longtable}
\usepackage{color}
\slugcomment{version \today: fm}

\shorttitle{Infrared colors of the $\gamma$-ray detected blazars}
\shortauthors{R. D'Abrusco, F. Massaro, M. Ajello, J. E. Grindlay, H. A. Smith, G. Tosti 2011}

\begin{document}
\title{Infrared colors of the Gamma-Ray detected Blazars}
\author{R. D'Abrusco\altaffilmark{1}, F. Massaro\altaffilmark{2}, M. Ajello\altaffilmark{2}, 
J. E. Grindlay\altaffilmark{1}, Howard A. Smith\altaffilmark{1} \& G. Tosti\altaffilmark{2,3,4}.}

\affil{Harvard - Smithsonian Astrophysical Observatory, 60 Garden Street, Cambridge, MA 02138}
\affil{SLAC National Laboratory and Kavli Institute for Particle Astrophysics and Cosmology, 2575 
	Sand Hill Road, Menlo Park, CA 94025}
\affil{Dipartimento di Fisica, Universita\` degli Studi di Perugia, 06123 Perugia, Italy}
\affil{Istituto Nazionale di Fisica Nucleare, Sezione di Perugia, 06123 Perugia, Italy}

\begin{abstract}

Blazars constitute the most enigmatic class of extragalactic $\gamma$-ray sources, and their observational 
features have been ascribed to a relativistic jet closely aligned to the line of sight. They are generally divided 
in two main classes: the BL Lac objects (BL Lacs) and the Flat Spectrum Radio Quasars (FSRQs). In the case 
of BL Lacs the double bumped spectral energy distribution (SED) is generally described by the 
Synchrotron Self Compton (SSC) emission, while for the FSRQs it is interpreted as due to External Compton (EC) emission.
Recently, we showed that in the [3.4]-[4.6]-[12] $\mu$m color-color diagram
the blazar population covers a distinct region (i.e., the \strip\,, WBS), clearly separated from the other extragalactic sources that are
dominated by thermal emission. In this paper we investigate the relation between the infrared and $\gamma$-ray emission 
for a subset of confirmed blazars from the literature, associated with \fer\ sources, for which \wse\ archival observations 
are available. This sample is a proper subset of the sample of sources used previously, and the availability of \fer\ data 
is critical to constrain the models on the emission mechanisms for the blazars. 
We found that the selected blazars also lie on the \strip\ covering a {\b narrower} region of the infrared color-color planes
than the overall blazars population. We then found an evident correlation between the IR and $\gamma$-ray spectral indices 
expected in the SSC and EC 
frameworks. Finally, we determined the ratio between their $\gamma$-ray and infrared fluxes, a surrogate of the ratio of powers
between the inverse Compton and the synchrotron SED components, and used such parameter to test different emitting 
scenarios blazars.
\end{abstract}

\section{Introduction}
\label{sec:intro}

Blazars are an intriguing class of active galactic nuclei (AGNs), dominated by non-thermal radiation over 
the entire electromagnetic spectrum. Their emission extends from radio to TeV energies with a broad band 
spectral energy distribution (SED) typically described by two main components, the first peaking from IR to 
X-ray bands while the second often dominating the $\gamma$-ray energy range in which blazars are the 
most commonly detected extragalactic sources \citep[e.g.][]{abdo10,massaro11a}.

The distinguishing observational properties of blazars also include flat radio spectra,
high observed luminosity coupled with rapid variability at all frequencies and highly 
variable radio to optical polarization. In particular, they are a dominant class of 
extragalactic sources at radio, microwave and $\gamma$-ray frequencies where 
thermal emission processes do not produce significant amounts of radiation \citep{giommi11}.
Adopting the blazar classification scheme given in the ROMA-BZCAT 
catalog\footnote{\underline{http://www.asdc.asi.it/bzcat/}} \citep{massaro09,massaro10},
based on the width of optical spectral lines, blazars usually come in two flavors: the 
BL Lac objects and Flat Spectrum Radio Quasars (FSRQs), with the latter more or less equivalent 
to the former population except for stronger emission lines, higher radio to optical polarization
and higher redshift. In addition, the high energy component of blazars SED 
is usually found in the GeV band for FSRQs, while it can extend to TeV energy for BL Lacs. Hereinafter, 
we adopt the naming convention of the ROMA-BZCAT catalog, referring to BL Lacs as BZBs and to 
the FSRQs as BZQs.

These extreme features have been interpreted as radiation arising from a relativistic 
jet closely aligned to the line of sight and emitting continuous, Doppler-boosted spectra \citep{blandford78}.
For both classes of blazars, according to the widely accepted Synchrotron Self Compton (SSC) scenario,
the low energy component is produced by inverse synchrotron emission from highly relativistic 
electrons, while the high energy bump can be attributed to inverse
Compton scattering of synchrotron photons by the same population of
relativistic electrons that produce the synchrotron emission \citep{marscher85,inoue96}. A different
theoretical interpretation of the high energy $\gamma$-ray bump characteristic of the BZQ broad band emission
invokes the inverse Compton emission of seed photons arising from regions external to the jet (e.g., 
the broad line region, accretion disk) as emission mechanism for such feature of the SED.
This model is known as the External Compton scenario (EC) \citep[e.g.,][]{dermer02,cavaliere02}.

Recently, we showed that in the [3.4]-[4.6]-[12] $\mu$m color-color diagram the blazars, 
which are dominated by non-thermal emission mechanism, cover a distinct region (hereinafter 
the \strip\,, WBS), well separated from the locus of other extragalactic sources which are dominated 
by the contribution of thermal radiation\citep{massaro11b}.

This result has been obtained using the blazars listed in the ROMA-BZCAT catalog. In the IR, we used data
from WISE \citep{wright10}. The \wse\ mission observed the sky at 3.4, 4.6, 12, and 22 $\mu$m 
in 2010 with an angular resolution of 6.1$^{\prime\prime}$, 6.4$^{\prime\prime}$, 6.5$^{\prime\prime}$ 
\& 12.0$^{\prime\prime}$ in the four bands, achieving 5$\sigma$ 
point source sensitivities of 0.08, 0.11, 1 and 6 mJy in unconfused regions on the ecliptic, respectively \citep{wright10}. 
The astrometric accuracy of \wse\ is $\sim 0.50^{\prime\prime}, 0.26^{\prime\prime}, 0.26^{\prime\prime}$, and 
1.4$^{\prime\prime}$ for the four \wse\ bands, respectively \citep{cutri11}.

A previous attempt to compare the infrared behavior of blazars with normal galaxies in the J-H-K color-color diagram
was performed using the 2MASS archival data \citep[e.g.,][]{chen05}.
However, our new approach has three advantages over the study performed using
2MASS: 1) mid-IR selection is dominated by dusty objects, in particular spiral and starburst galaxies; 2) 
the blazars population covers a noticeably narrow region in the [3.4]-[4.6]-[12] $\mu$m 
color-color plot that is clearly statistically separated them from the locus dominated by 
other extragalactic sources \citep{massaro11b}; 
3) \wse\ covers a much larger interval of frequencies and reaches to larger wavelengths than 2MASS, 
yielding a color-color distribution where stars occupy a narrow and well defined locus.

The \fer\ Large Area Telescope (LAT) collaboration has recently released the \fer\ LAT second source catalog 
(2FGL) \citep[e.g.][]{abdo11} including about 800 $\gamma$-ray sources associated with blazars to a high level of 
confidence; 571 of these blazars are also present in the ROMA-BZCAT.
In this paper we investigate how the infrared emission of blazars is related to their high energy 
$\gamma$-ray radiation, by cross-matching the sample of 
ROMA-BZCAT blazars associated with \fer\ sources with the \wse\ data archive \citep{wright10}. 
In Section~\ref{sec:sample}, we present the blazar sample used throughout our investigation.
In Section~\ref{sec:2FBS} we show how the \fer\ detected blazars lie on the \strip\ and we determine the 
distributions of their IR colors. Then, in Section~\ref{sec:correlation}, we investigate a possible relation 
between the blazar IR and $\gamma$-ray emissions, as predicted in the SSC and EC radiative scenarios. 
In Section~\ref{sec:cd} we estimate the Compton Dominance (CD) parameter for both the BZQs and the BZBs and 
we also examine the apparent correlation between the CD and the $\gamma$ -ray spectral index for BZBs. 
In Section~\ref{sec:bllacs} we focus our analysis on the BL Lac population and its subclasses.
Finally, our summary and discussion are given in Section~\ref{sec:summary}.

Throughout this paper, we assume that the spectral indices $\alpha$, are defined by flux density, 
$S_{\nu}\!\propto\!\nu^{-\alpha}$. Unless otherwise stated, we use cgs units.

\section{Sample Selection}
\label{sec:sample}

We considered all the blazars in the ROMA-BZCAT that have been associated with a \fer\ source, as reported in the 
2FGL \citep{abdo11}, for a total number of 571 sources (i.e., 330 BZBs and 241 BZQs).
The second edition of the ROMA-BZCAT catalog \citep{massaro09} assembles blazars known in the 
literature and carefully verified by inspection of their multiwavelength emission. Members of the ROMA-BZCAT catalog
are selected on the basis of a set of criteria involving the presence of detection in the radio band 
down to 1 mJy flux density at 1.4 GHz {\bf (2.1 $\mu m$)}, the optical identification and availability of an optical spectrum for 
further spectral classification, and the detection of X-ray luminosity $L_X \geq 10^{43}$ ergs$\cdot s^{-1}$.
Such criteria do not produce a statistically homogeneous neither complete sample of blazars because of the 
spatially uneven distribution and the variable depths of observations available, but provides the largest and more 
carefully selected sample of confirmed blazars available to date.  
The selection and spectral classification of blazars can be difficult due to the absence of typical spectral emission 
features and to the variability of the emission on timescales of a day or even 
few hours. In the ROMA-BZCAT, blazars are also classified in three classes, based on prominence of the emission 
features in the optical spectra of these sources. The three classes of such classification are: BZB for the BL Lac sources, i.e. AGNs 
with featureless optical spectra and narrow emission lines; BZQ for flat-spectrum radio quasars with optical spectra 
showing broad emission lines and 
typical blazars behavior; BZU for blazars of uncertain type, associated to sources with peculiar characteristics but also 
showing typical traits of the blazars. This spectral classification will be used throughout this paper. The 
distinction between BZB and BZQ depends on the choice of an arbitrary threshold value of the equivalent width
of the emission lines in the optical spectra of the sources.

The 2FLG catalog contains primarily unresolved sources detected in the all-sky Fermi observations obtained 
throughout the second year of operation. The sources, after detection and the localization in the sky, are assigned
an integrated flux in the 100 MeV to 100 GeV energy range, a spectral shape and a significance parameter {\it TS} 
based on how significantly each source emerges from the background. Only sources with
{\it TS $\geq$ 25}, corresponding to a significance of $4\sigma$, have been included in the catalog. Each of the 1873 
2FLG sources have been considered for identification with already known astronomical sources available in 
literature multi-wavelength observations \citep{abdo11}. For 127 of the 2FLG  sources firm identifications have 
been produced (namely, reliable identifications based on synchronous periodic variability of the sources, 
coincident spatial morphologies for extended sources or correlated aperiodic variability). The remaining
sources have been investigated for association with sources contained in a list of source catalogs based on 
different multi-wavelength observations. The ROMA-BZCAT catalog is one of the catalogs
used for the association of the 2FLG sources, and 571 \fer\ sources have been associated to
a BZCAT-ROMA blazar. 

We selected all blazars in the ROMA-BZCAT reliably associated to a $\gamma$-ray 
source of the 2FGL catalog. Then, using the more accurate radio position of the ROMA-BZCAT in place 
of the coordinates from the \fer\ catalog, we searched for infrared counterparts of the above blazars in 
the \wse\ archive.

The total number of ROMA-BZCAT blazars in the 2FGL footprint falling in the area surveyed by 
\wse\ during the first year (corresponding to 57\% of the whole sky) is 332
(mostly due to the incompleteness of the sky coverage of the ROMA-BZCAT). 
In order to search for the positional coincidences of blazars in the observed \wse\ sky,
we considered a search radius 2.4$^{\prime\prime}$, obtained by combining the 1$^{\prime\prime}$ error 
assumed for the radio position reported in the ROMA-BZCAT \citep{massaro09} with the error on the fourth 
\wse\ band at 22$\mu$m (i.e., 1.4$^{\prime\prime}$) \citep[see also][, for more details]{wright10}.
Using a conservative approach in our analysis, we only considered sources in \wse\ Preliminary Source Catalog (WPSC) 
\footnote{\underline{http://wise2.ipac.caltech.edu/docs/release/prelim/preview.html}} with a minimum signal-to-noise ratio 
(SNR) of 7 in all the four infrared bands.

The number of \fer\-BZCAT blazars with a \wse\ counterpart within the first region of 2.4$^{\prime\prime}$ is 296,
corresponding to $\sim$45\% of the \fer\ - \wse\ blazars sample in the WPSC, detected
with a chance probability of $\sim$3\% \citep[see][for more details]{maselli11,massaro11b}
We did not find any multiple matches using 2.4$^{\prime\prime}$ as search radius.
We have used only these 296 blazars for our investigation. The remaining 12 sources 
with no apparent corresponding \wse\ sources within 2.4$^{\prime\prime}$ can be associated to 
at least one source contained in the \wse\ catalog using a search radius of 12$^{\prime\prime}$, but
have not been used in our study. By definition, 
the \fer\ - \wse\ blazars sample is a proper subset of the sample of ROMA-BZCat blazars with \wse\ 
counterparts discussed in the previous paper \citep{massaro11b}. The table containing the main 
parameters of the 296 sources of the \fer\ - \wse\ blazars sample is published in the Appendix 
(Section~\ref{sec:appendix2}).

\section{The \fer\ blazars at infrared frequencies}
\label{sec:2FBS}

\begin{figure}
\includegraphics[height=7.5cm,width=8.5cm,angle=0]{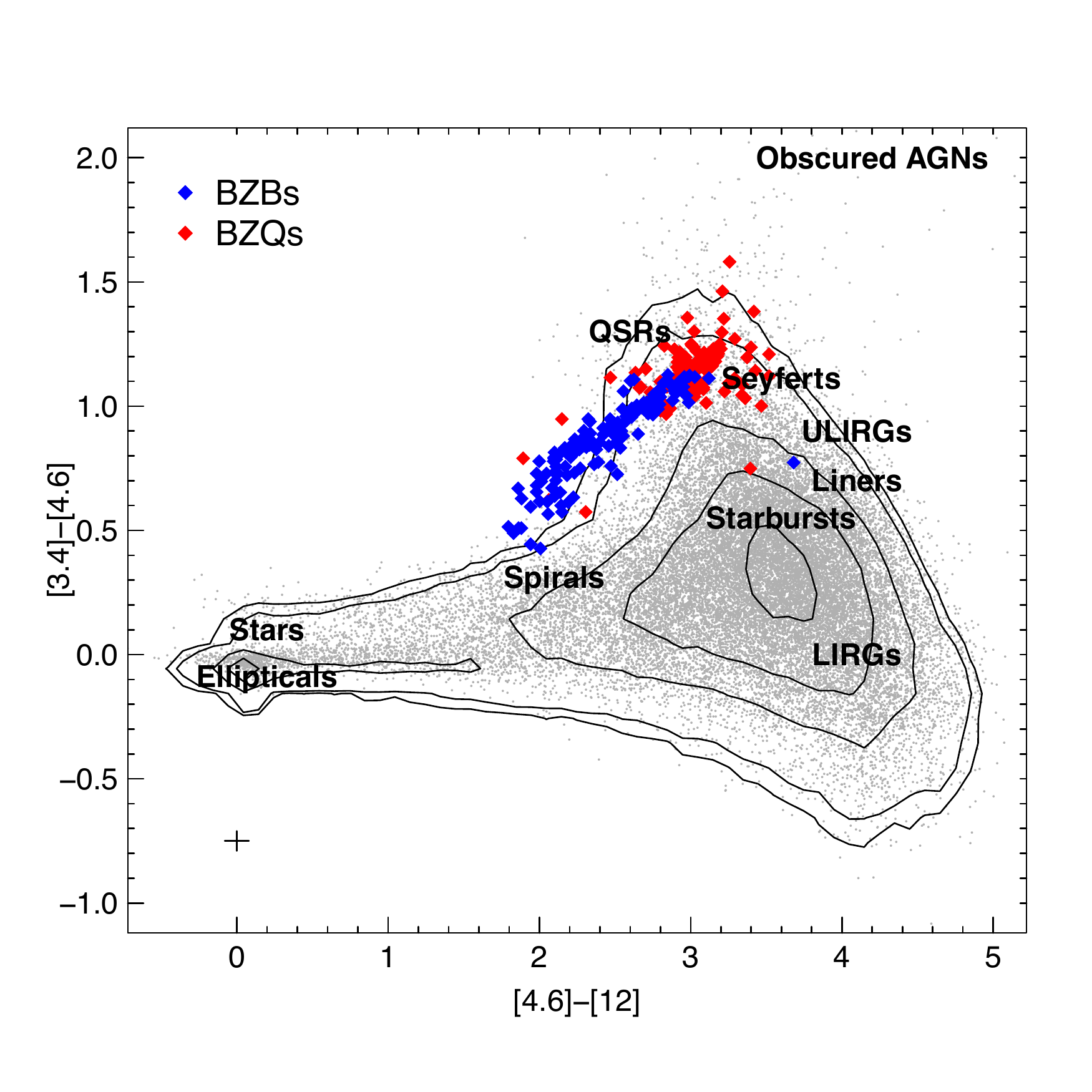}
\caption{The [3.4]-[4.6]-[12] $\mu$m color-color diagram of \fer\ - \wse\ blazars sample sources.
We plot the 296 blazars associated with a \wse\ source within a region of radius 2.4$^{\prime\prime}$.
The two blazar classes of BZBs (blue) and BZQs (red) are shown. The background grey dots correspond 
to 453420 \wse\ sources detected in a region of 56 deg$^2$ at high Galactic latitude. The isodensity 
curves for the \wse\ sources, corresponding to 50, 100, 500, 2000 sources per unit area in the color-color 
plane respectively, are shown (see Section~\ref{sec:2FBS}). The location of different classes of objects 
is also shown, with QSRs, ULIRGs and LIRGs indicating the quasars, the ultraluminous infrared galaxies 
and the luminous infrared galaxies respectively.}
\label{fig:color2color1}
\end{figure}

As recently shown in \citep{massaro11b}, the blazars, dominated in the infrared by their synchrotron emission,
lie in a distinct region of the [3.4]-[4.6]-[12] $\mu$m color-color diagram, and appear to be distinctly separated 
from the rest of the not-synchrotron dominated sources populating the sky as observed by \wse\ . 
We selected 14 not-overlapping, randomly picked regions of 4 deg$^2$ each for a total 56 deg$^2$ at high 
Galactic latitude \citep{massaro11b}, within the 116 
deg$^2$ considered in the WSPC \citep{cutri11}. We collected all the 453420 sources detected by \wse\ in its 1st year catalog 
(hereinafter called \wse\ thermal sources because most of them are dominated by thermal emission in the infrared energy range),
having SNR$>$7 in at least one band, a conservative level for the WPSC release to emphasize the catalog reliability$^{3}$ 
\citep{cutri11}. We have not excluded the stars from the sample of generic \wse\ detected sources in the color-color
diagrams shown in this paper, since at high galactic latitude, the majority of the observed sources are extragalactic with 
only little contamination from stars, and we have checked their presence do not negatively affect the conclusions about 
the separation of the region of color space occupied by \fer\ detected blazars.

We built the [3.4]-[4.6]-[12] $\mu$m color-color diagrams from the magnitudes reported in the \wse\ 
catalog\footnote{All \wse\ magnitudes are in Vega system.} for all the \wse\ thermal sources in the 
56 deg$^2$ area described above, and for all the sources in the \fer\ - \wse\ blazars sample.
In Fig.~\ref{fig:color2color1} we also show the location of different classes of objects, and 
overlaid to five levels isodensity contours for all the \wse\ thermal sources in the 56 deg$^2$ region.
We plot the blazars of the diagram to characterize their infrared emission.
Fig.~\ref{fig:color2color1} shows that the \fer\ - \wse\ blazars lie in an even more confined region than the general
\strip\ shown in Fig. 1 of \citep{massaro11b}.
We note that the relative errors for both the infrared colors are less than 10\%
for 97\% of the \fer\ - \wse\ blazars sample but less than 5\% for $\sim$ 85\% of the sources.

The subregion of the \strip\ occupied by the \fer\ - \wse\ blazars is well defined. 
Only two ``outliers" out of a total number of 296 blazars are visible in this color-color plane: 2FGL J1506.6+0806 and 
2FGL J1550.7+0526 (also known as 4C\,+05.64 or PKS\,1548+056).
While for these two specific sources the possibility of a wrong association in the 2FGL catalog cannot be ruled
out, in general a possible explanation for these or other ROMA-BZCAT sources to lie outside of the \strip\ 
might be a thermal contribution from their host galaxy that is non-negligible with respect to the 
non-thermal IR emission \citep[e.g.][]{massaro11d}. In general, however, the blazars of the \fer\ - \wse\ blazars 
sample, dominated by synchrotron emission in the IR are located in distinctly defined regions of the 
\wse\ color-color planes all well separated from the other non synchrotron dominated sources detected
by \wse\ .

\begin{figure}[!b]
\includegraphics[height=6.2cm,width=8.5cm,angle=0]{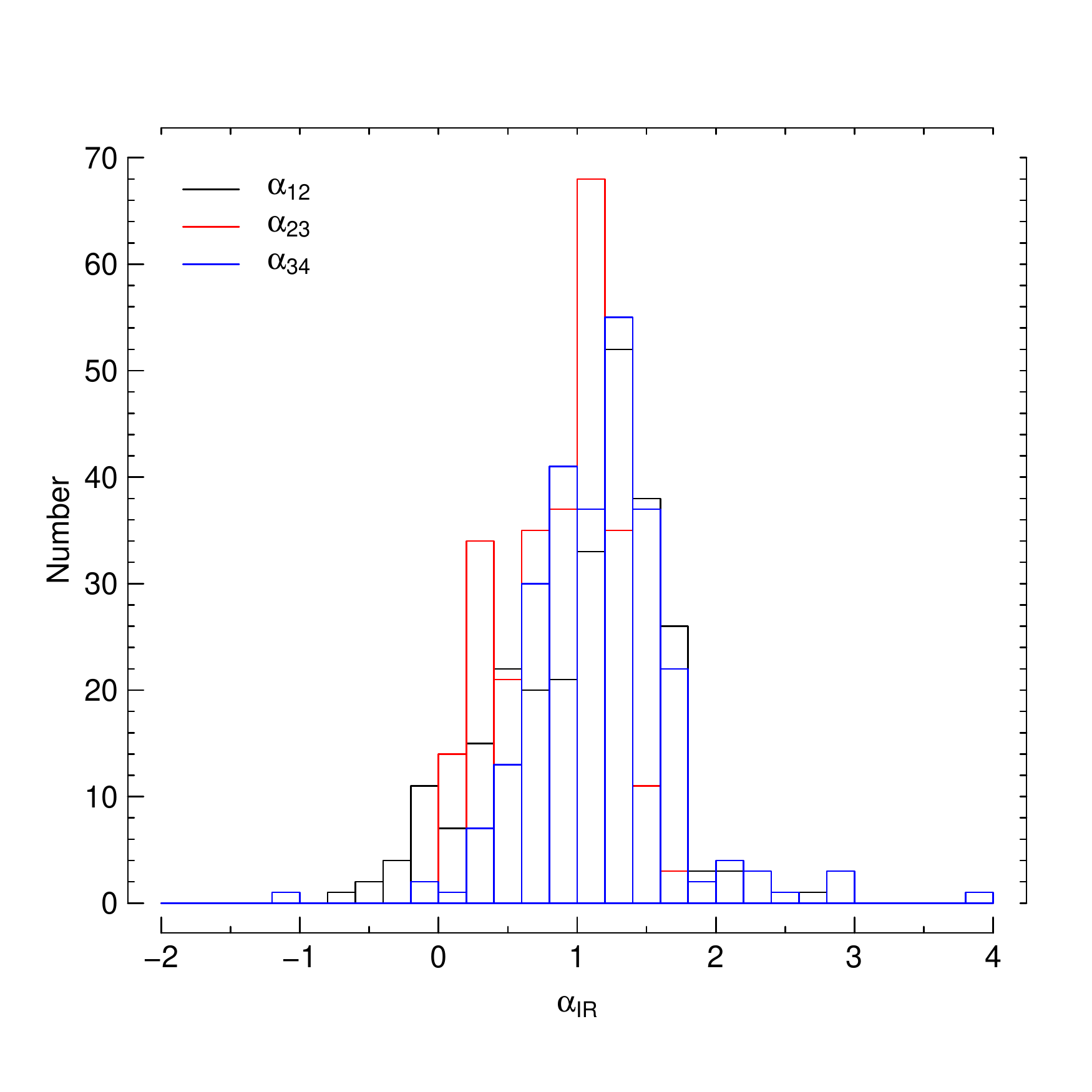}
\caption{Histograms of the distributions of the infrared spectral indices $\alpha_{12}$, $\alpha_{23}$, $\alpha_{34}$
for the \fer\ - \wse\ blazars sample, derived by using Equation~\ref{eq:color} for the three different colors (see 
Section~\ref{sec:2FBS} for more details).}
\label{fig:histoinfraredspectralindexes}
\end{figure}

Assuming that the infrared spectrum of the sources in the \fer\ - \wse\ blazars sample can be 
described by a power-law, we derive the relation between the infrared colors and the spectral slope $\alpha$.
Considering a source of apparent magnitudes $m_1$, $m_2$, $m_3$, $m_4$ in the four different \wse\ bands, with the
zero-point magnitudes $m_{01}$, $m_{02}$, $m_{03}$ and $m_{04}$ respectively, the relation between one color,
for example $c_{12}=m_1-m_2=[3.4]-[4.6]$ and the associated spectral slope $\alpha_{12}$ can be written as:

\begin{equation}
c_{12} = m_1-m_2 = 2.5\,\alpha_{12}\,\log\left(\frac{\nu_1}{\nu_2}\right)\,+\,(m_{01}-m_{02})
\label{eq:color}
\end{equation}

\noindent where $\nu_1$ and $\nu_2$ are the frequencies corresponding to the 3.4 and 4.6$\mu$m wavelengths
respectively. We estimated the values of the spectral indices $\alpha_{12}$, $\alpha_{23}$, $\alpha_{34}$ from the 
three infrared colors $c_{12}$, $c_{23}$ and $c_{34}$, respectively.
Then we compared their distributions to test for the presence of spectral curvature for the sources in the 
\fer\ - \wse\ blazars sample.
We found that the median (1.07, 0.94 and 1.12 for $\alpha_{12}$, $\alpha_{23}$ and $\alpha_{34}$), the peak values
(1.06, 0.99 and 1.12) and variances (0.33, 0.22 and 0.26) of the distributions of the three IR spectral indices are 
consistent with each other. In the following analysis we will consider the infrared spectral 
index $\alpha_{IR}=\alpha_{12}$, because the \wse\ 3.4 and 4.6 $\mu$m filters are the most sensitive.
Unfortunately, given the \wse\ restricted energy range we did not find any hint of a curved spectral shape.
In Fig.~\ref{fig:histoinfraredspectralindexes}, we show the distribution of the spectral indices 
in each band, $\alpha_{12}$, $\alpha_{23}$, $\alpha_{34}$.

\begin{figure}
\includegraphics[height=7.5cm,width=8.5cm,angle=0]{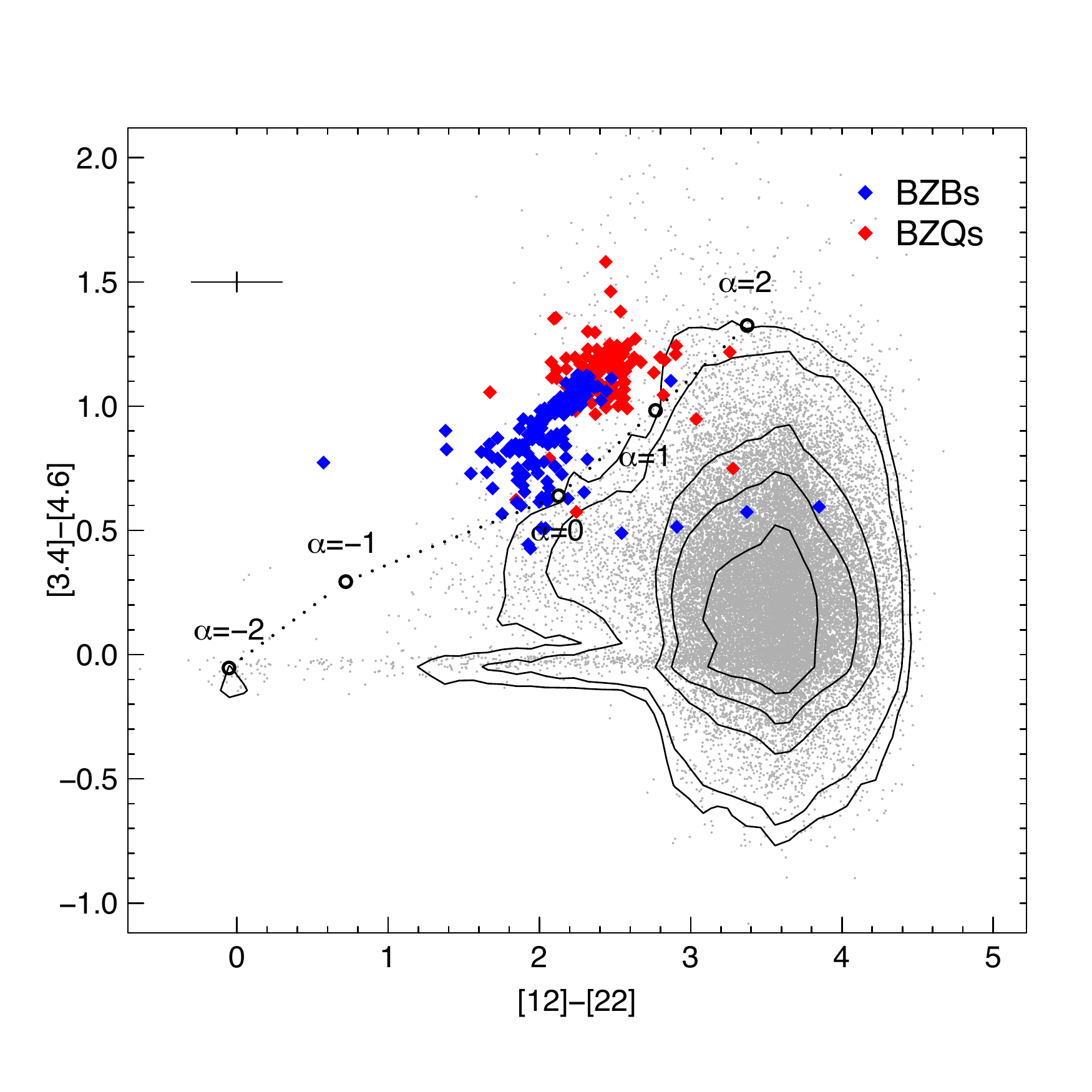}
\caption{Same of Fig.~\ref{fig:color2color1} with a different choice of infrared colors: [3.4]-[4.6]-[12]-[22] $\mu$m.
We also report the black dashed line corresponding to the IR colors generated by a power-law spectrum
of spectral index $\alpha_{IR}$. The black cross shown in the right bottom represents the typical error 
bars on the infrared colors (see Section~\ref{sec:2FBS} for more details).}
\label{fig:color1color3}
\end{figure}

Since the sources in the \fer\ - \wse\ blazars sample have been detected in all the four \wse\ infrared bands,
we can also construct the [3.4]-[4.6]-[12]-[22] $\mu$m color-color diagram, where the two colors are independent 
(see Fig.~\ref{fig:color1color3}). In this color-color diagram, the separation between the blazars and the generic
\wse\ sources is even more evident than in the [4.6]-[12]-[22] $\mu$m color-color diagram, even though the locus is 
less narrow than in the case of the [3.4]-[4.6]-[12]-[22] color-color diagram shown in Fig.~\ref{fig:color2color1}). 
In this plot we {\bf also} report the line corresponding to a power-law spectrum of {\bf varying} indices $\alpha_{IR}$ 
described by the equation:

\begin{equation}
\label{eq:retta}
c_{12} = \left[\frac{\log(\nu_1/\nu_2)}{\log(\nu_3/\nu_4)}(c_{34}-m_{03}+m_{04})\right]+(m_{01}-m_{02}) 
\end{equation}

\noindent where $c_{34}=m_3-m_4=[12]-[22]$ is the infrared color corresponding to the bands at 12$\mu$m and 
22$\mu$m respectively (see Fig.~\ref{fig:color1color3}).

\noindent The remaining [4.6]-[12]-[22] $\mu$m color-color diagram is shown in the appendix (Section~\ref{sec:appendix}.)
In particular, we note that in Fig.~\ref{fig:color1color3} (and Fig.~\ref{fig:color2color3} in the 
Appendix~\ref{sec:appendix}), the regions covered by the \fer\ - \wse\ blazars sample sources are also clearly 
separated from the thermal \wse\ sources.

\begin{figure}
\includegraphics[height=7.5cm,width=8.5cm,angle=0]{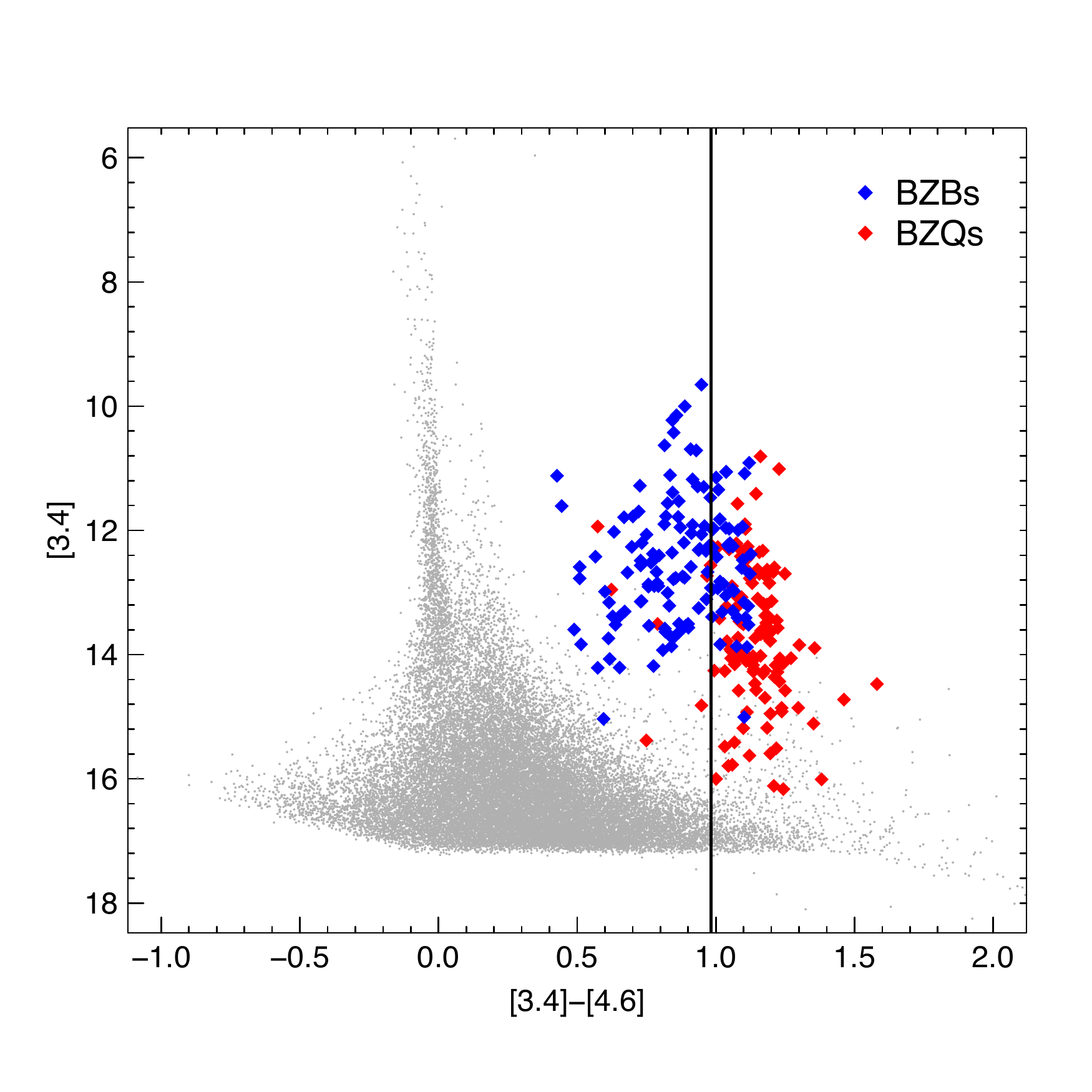}
\caption{The color-magnitude diagram built with the most sensitive infrared \wse\ bands, {namely the [3.4]-[4.6] color 
and the [3.4] magnitude}. The BZQs sources clearly show value of the color [3.4]-[4.6] $\sim$ 1 corresponding to a 
spectral index $\alpha_{IR}$ of -1, as a consequence of the \wse\ observations sampling the peak of their 
synchrotron components. The black vertical line corresponds to the color value associated to a power law with 
spectral index $\alpha_{\nu}=-1$.}
\label{fig:color1mag1}
\end{figure}

Finally, we present a color-magnitude diagram for the three \wse\ bands with highest sensitivity,
(Fig.~\ref{fig:color1mag1}), where the flux limit of the \wse\ survey 
is clearly visible. In this plot all \fer\ - \wse\ blazars sample sources lie well above the value of the 
limiting magnitude at 3.4$\mu$m, and the blazars appear significantly brighter than all the other 
sources with similar values of the color. Nonetheless, as already discusses for  [3.4]-[4.6]-[12]-[22] 
color-color diagram shown in Fig.\ref{fig:color2color1}, if on one hand the sources of the \fer\ - \wse\ blazars 
sample are well separated from the \wse\
sources even in this color-magnitude plot, on the other hand the region of the plane occupied by the 
blazars is less compact and well defined than the \strip\ visible in the [4.6]-[12] $\mu$m vs [3.4]-[4.6] $\mu$s 
color-color plane (Fig.~\ref{fig:color2color1}). Other color-magnitude plots of the \fer\ - \wse\ blazars sample 
are shown in the Appendix (Section~\ref{sec:appendix}).

\noindent It is worth stressing that $\sim 95\%$ of the BZQs have [3.4]-[4.6] color larger than the value 
of the color associated with a power law spectrum of spectral index 1. This fact suggests that the peaks of the 
first component (i.e., the synchrotron emission) of these sources occurs inside or very close to the \wse\ spectral range. 
The situation appear to be different for the BZB class, which displays infrared colors ranging between 0.5 and 1.2 (see 
Section~\ref{sec:bllacs} for more details). It is also interesting to note that all the sources in the \fer\ - \wse\ blazars sample 
are consistently above the sensitivity limit of the \wse\ survey, even though this effect is most likely due to the luminosity 
distribution and selection limits of the \fer\ observations.

\section{The $\alpha_{IR}-\alpha_{\gamma}$ correlation}
\label{sec:correlation}

According to the SSC or the EC scenarios, usually adopted to interpret the blazars emission, 
the particles (i.e., electrons) that are emitting via synchrotron radiation at
radio and infrared frequencies are also those that are scattering the photons to 
high energy, in the X-rays and in the $\gamma$-rays, via inverse Compton emission.
Consequently, an empirical correlation between the spectral indices and the fluxes in the infrared and in the 
$\gamma$-ray energy range is expected, since they originate from the same electron distribution.

The relationships between radio, microwave and $\gamma$-ray emissions of blazars is discussed by many 
authors \citep[e.g.][]{giommi11}. A positive correlation between the radio and 
$\gamma$-ray fluxes has been observed, though with large scatter, using several samples 
\citep[e.g.,][]{kovalev09,giroletti10,leon11}. 
However, a correlation between $\gamma$-ray and IR in the spectral range covered by \wse\ had
not been observed to date. This could be mainly due to the lack of $\gamma$-ray observations for a 
sufficiently large number of blazars, now resolved by the availability of \fer\ data. 
At the same time, the infrared frequencies observed by \wse\ had not been extensively investigated for 
blazars despite the fact that the SED frequency peak lies very close or inside this spectral range, at 
least for most of the BZQs.

\begin{figure}
\includegraphics[height=7.5cm,width=8.5cm,angle=0]{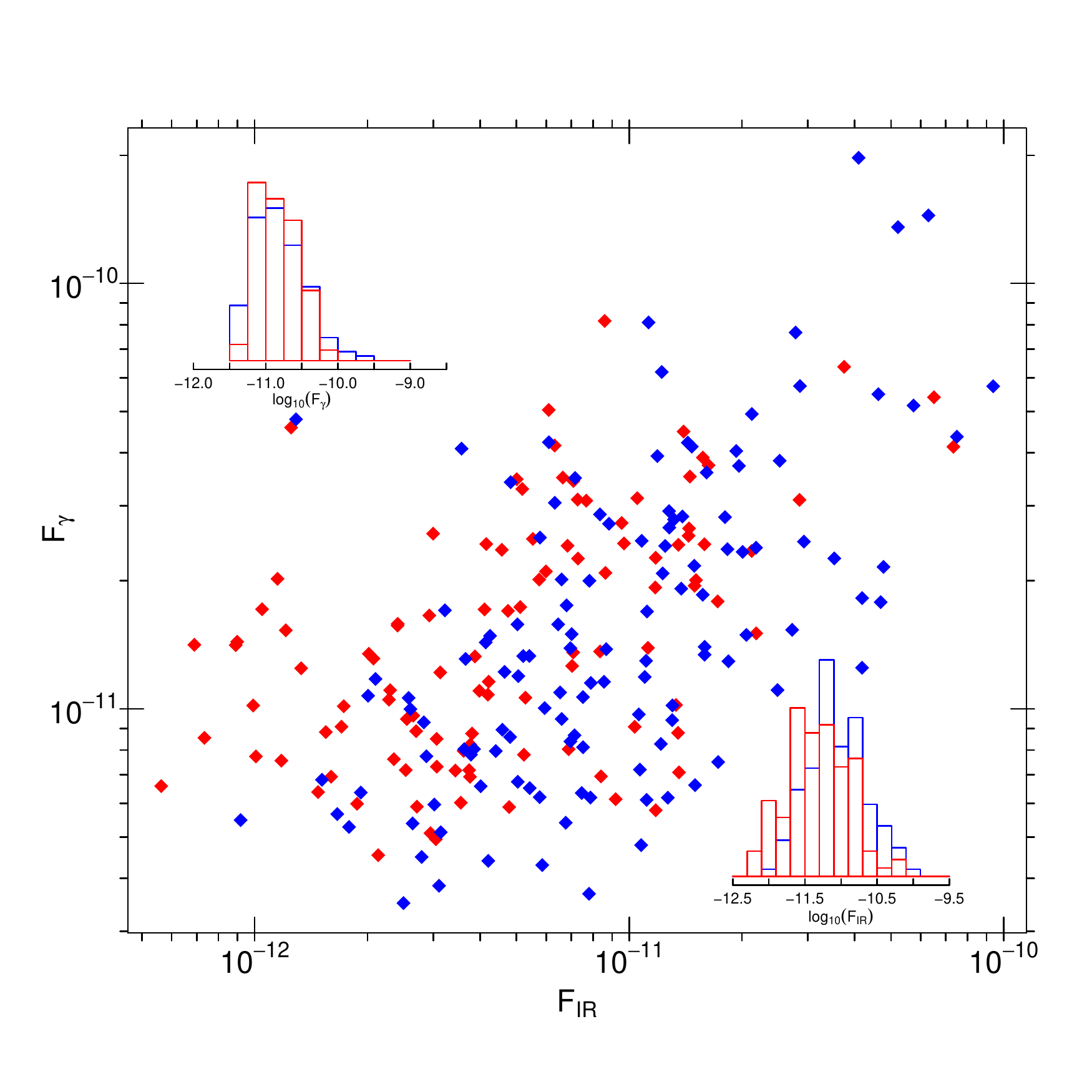}
\caption{Distribution of the sample of 2FGL detected blazars from the 
BZCAT-ROMA catalog with \wse\ counterparts relative to the total $\gamma$-ray flux
(from the 2FGL catalog) and the total IR \wse\ flux calculated from the magnitudes in the 
4 \wse\ filters. BZBs and BZQs according to the spectral classification available 
in the ROMA-BZCAT catalog are plotted with red and blue respectively. The histograms of the 
IR and $\gamma$-day fluxes distributions for the sample of blazars considered are shown in 
the two insets of the plot. All fluxes are expressed in cgs units.}
\label{fig:fluxflux}
\end{figure}

In Fig.~\ref{fig:fluxflux} we show the distribution of the sample of blazars considered in this paper 
in the plane of the total \fer\ energy flux in the 100MeV - 100 GeV 
energy interval obtained by spectral fitting in the same range (from the 2FGL catalog), and of the total 
\wse\ IR flux derived from the \wse\ magnitudes. 
The total IR fluxes have been calculated by summing the fluxes in the four WISE filter obtained 
as $\nu f(\nu)$ and accounting gor the color corrections as discussed in ~\citep{wright2010}.
BZBs and BZQs are plotted with different colors in this plot. A correlation between $\gamma$-ray 
and IR fluxes is statistically 
significant for the whole sample of sources (with correlation coefficient $r_s = 0.57$) and for 
both classes of BZBs and BZQs separately ($r_s = 0.64$ and $r_s = 0.61$ respectively). 

\begin{figure}
\includegraphics[height=7.5cm,width=8.5cm,angle=0]{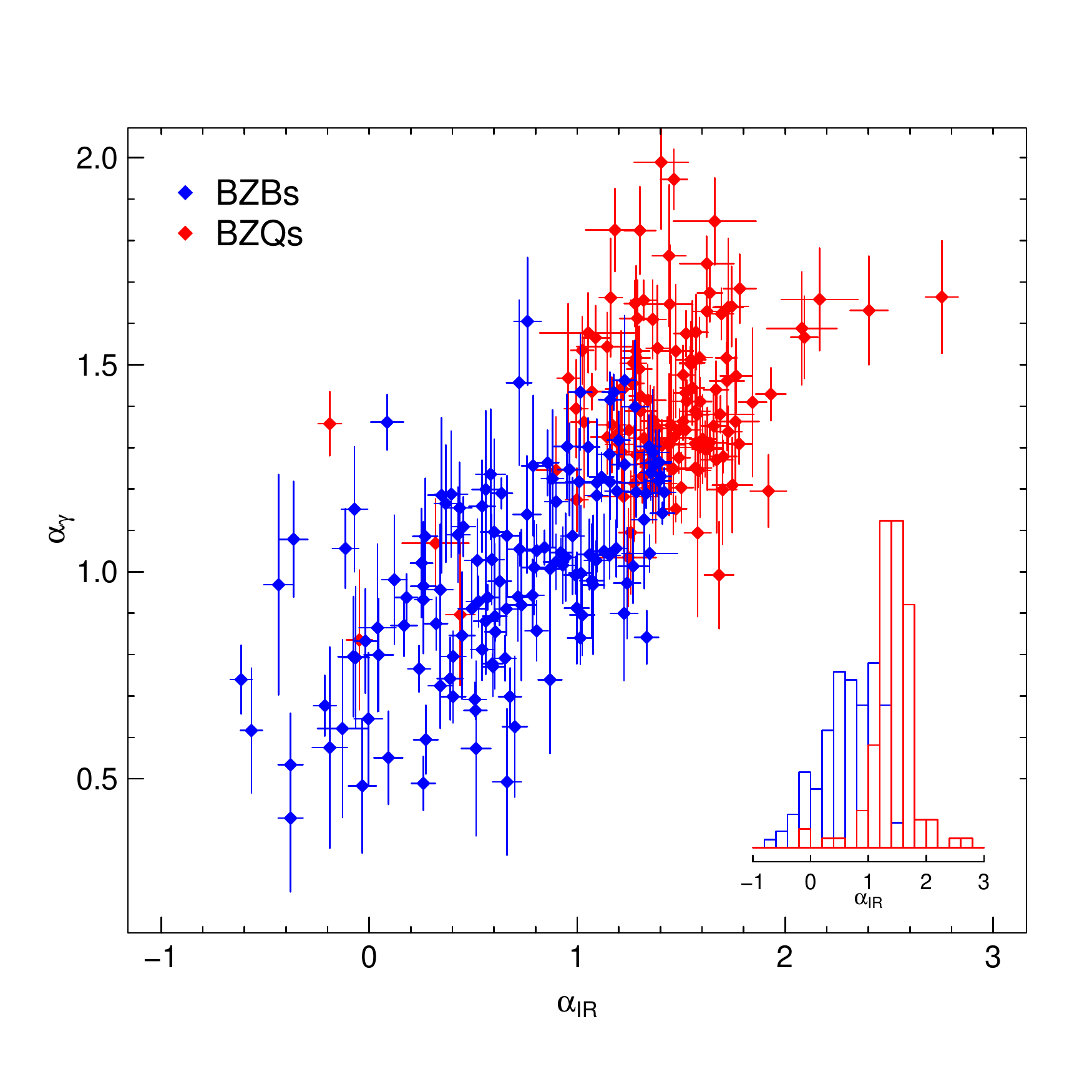}
\caption{Scatterplot of the $\alpha_{IR}-\alpha_{\gamma}$ distribution, with the two classes of blazars 
BZBs (blue) and BZQs (red) shown. In the inset, the histogram of the two distributions of $\alpha_{IR}$
is plotted, clearly showing the dichotomy between the two classes (see Section~\ref{sec:correlation} 
for more details). Three linear regression lines with different colors are shown: the black has been 
obtained by linear fitting on the whole sample of BZBs and BZQs, while the blue and red lines are 
associated to the regression evaluated for BZBs and BZQs alone, respectively.}
\label{fig:spectralindexIRspectralindexGAMMA}
\end{figure}

\begin{figure}
\includegraphics[height=7.5cm,width=8.5cm,angle=0]{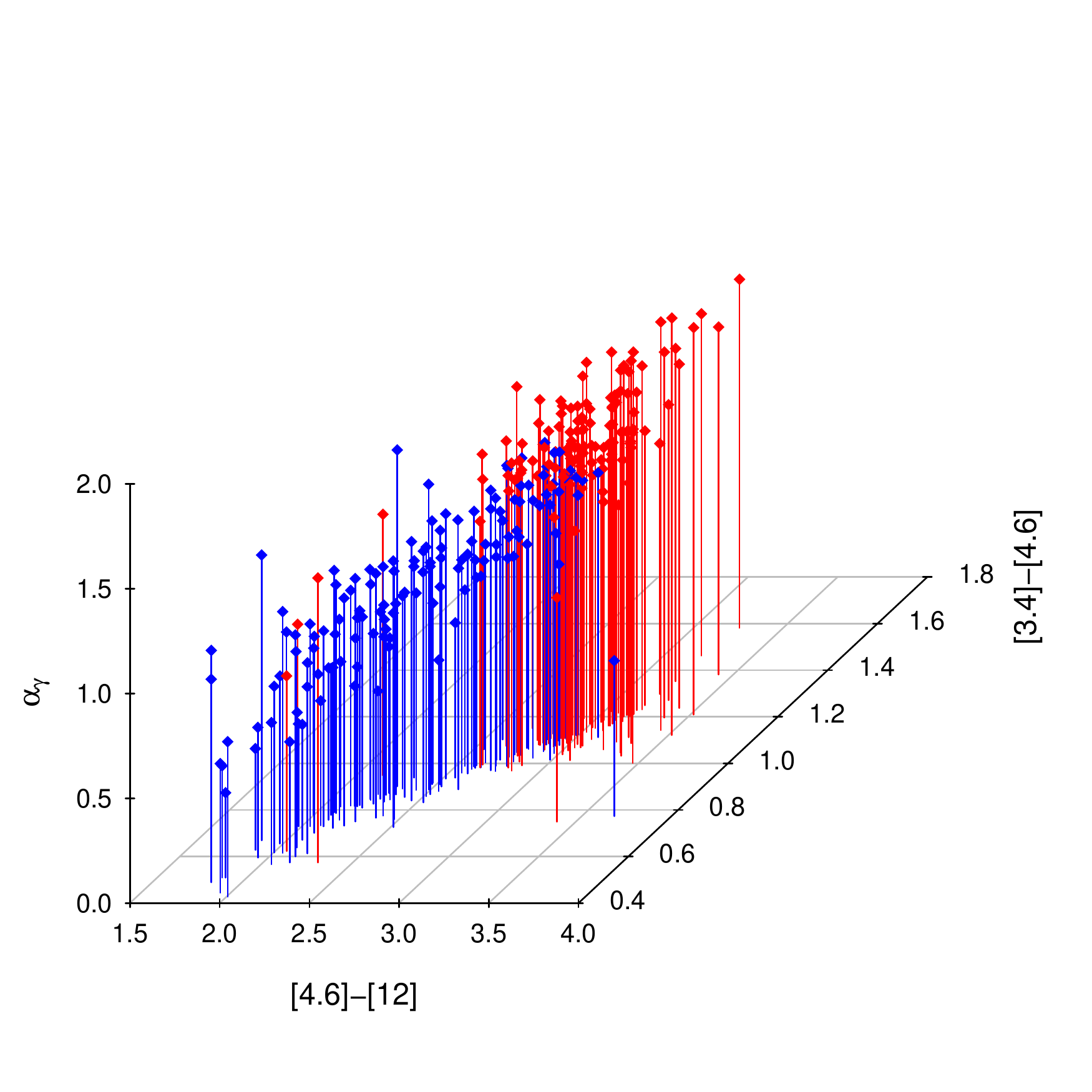}
\caption{Three dimensional scatterplot of the distribution of \fer\ blazars in the [4.6]-[12] - [3.4]-[4.6] - 
$\alpha_{\gamma}$ space. The two different classes of blazars, namely the BZBs (blue points) 
and BZQs (red points), are shown (see Section~\ref{sec:correlation} for more details). The lines
extending downwards are meant to show the position of the projections of the points on the [3.4]-[46] vs
[4.6]-[12] color-color plane.}
\label{fig:3Dscatterplot}
\end{figure}

In Fig.~\ref{fig:spectralindexIRspectralindexGAMMA} we report the $\alpha_{IR}-\alpha_{\gamma}$ 
scatterplot with the black line obtained by linear regression on the two spectral indices for
BZBs and BZQs together. The associated correlation coefficient is $r_s = 0.71$, corresponding to 
a negligible; this implies that the two spectral indices are correlated at a very high level of significance.
We have also evaluated the best fitting linear relations for BZBs and BZQs separately (blue and 
red lines respectively) and the Spearman's correlation coefficients for these two subsamples of sources. 
While the correlation between the two spectral indices for the BZBs has a high level of significance with 
$r_s = 0.59$ and negligible p-value, the correlation for the BZQs is associated to a lower values of
the correlation coefficient $r_s = 0.14$, so that the hypothesis that the two parameters are uncorrelated
can be rejected at a 80\% level of significance. Overall, the correlation between the IR and $\gamma$-ray
spectral indices is dominated by the BZBs.
The dichotomy between the BZB and BZQ classes of objects is evident from this plot, not only
in the $\alpha_{\gamma}$ distribution and in the comparison of the linear regression lines for the 
two classes of sources separated \citep[e.g.,]{abdo11}, but also in the $\alpha_{IR}$ distribution
(see the histogram in Fig.~\ref{fig:spectralindexIRspectralindexGAMMA}).
We performed a KS test and we found that the distributions of the $\alpha_{IR}$ for the 
BZBs and the BZQs differ at 97\% level of significance.
We also report in Fig.~\ref{fig:3Dscatterplot} the 3-dimensional plot of the two main infrared colors
used to build the [3.4]-[4.6]-[12] $\mu$m color-color diagram and the $\gamma$-ray spectral index
$\alpha_{\gamma}$, to highlight the distinction between the two classes of blazars.
The two colors can be here considered as surrogates of the infrared spectral index $\alpha_{IR}$
(see Section~\ref{sec:2FBS} for more details).

\section{The Compton dominance}
\label{sec:cd}

The Compton Dominance (CD) parameter, defined as the ratio between the
inverse-Compton and synchrotron peak luminosities, can be used to 
identify which is the main radiative loss process for the emitting particles, either
synchrotron or inverse Compton emission.
In particular,  since for the majority of the BZQs the \wse\ and the \fer\ spectral ranges 
are directly sampling or are very close to the peak frequency for both the inverse 
Compton and the synchrotron components, the ratio between the infrared and the 
$\gamma$-ray fluxes provides a good estimate of the CD parameter for the \fer\ - \wse\ 
blazars sample sources.

\begin{figure}[!b]
\includegraphics[height=6.2cm,width=8.5cm,angle=0]{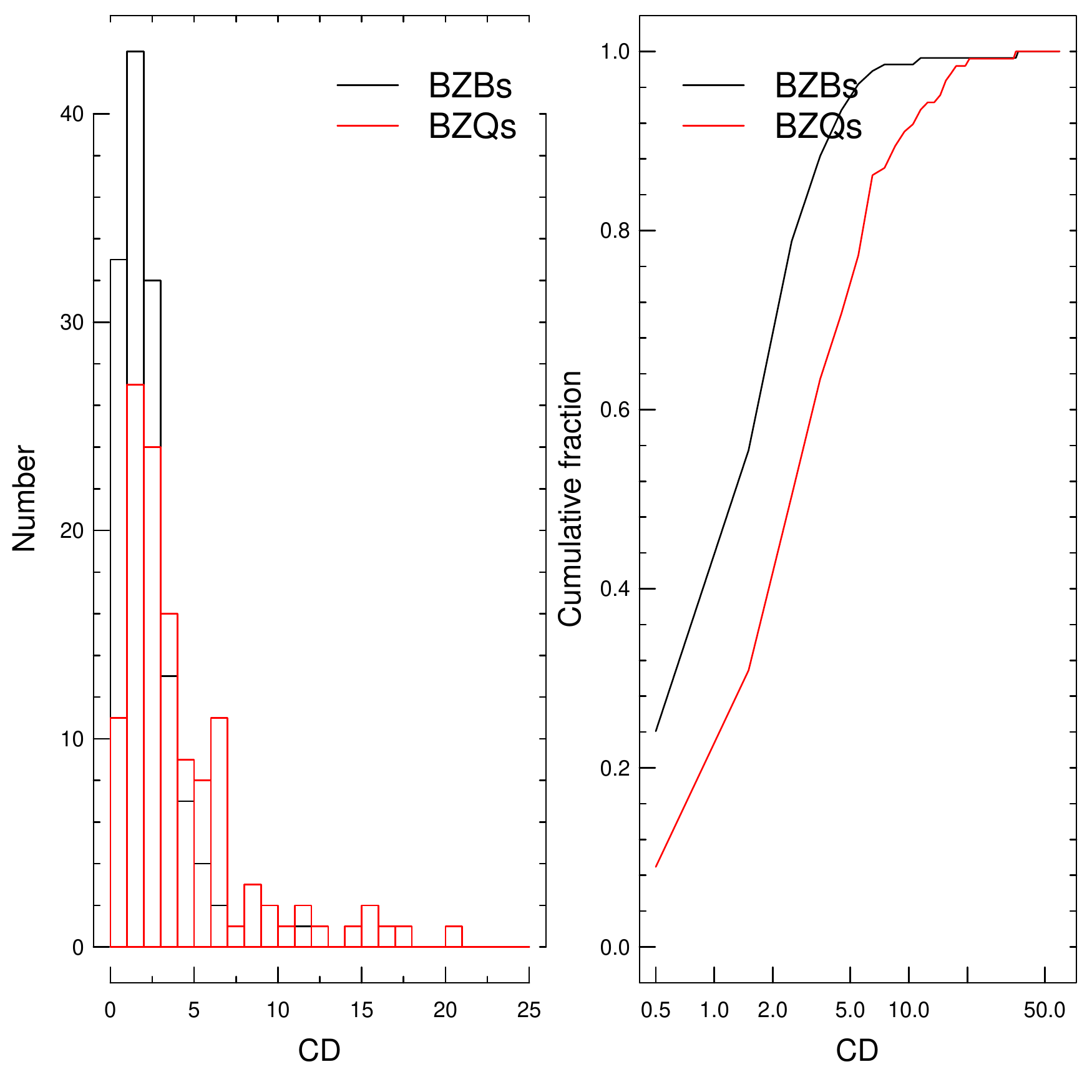}
\caption{Histogram of the distributions of the CD parameter for the BZBs (black) and the BZQs (blue) 
(left panel), and their cumulative distribution (right panel) (see Section~\ref{sec:cd} for more details).}
\label{fig:histocomptondominancecumulative}
\end{figure}

\begin{figure}
\includegraphics[height=6.2cm,width=8.5cm,angle=0]{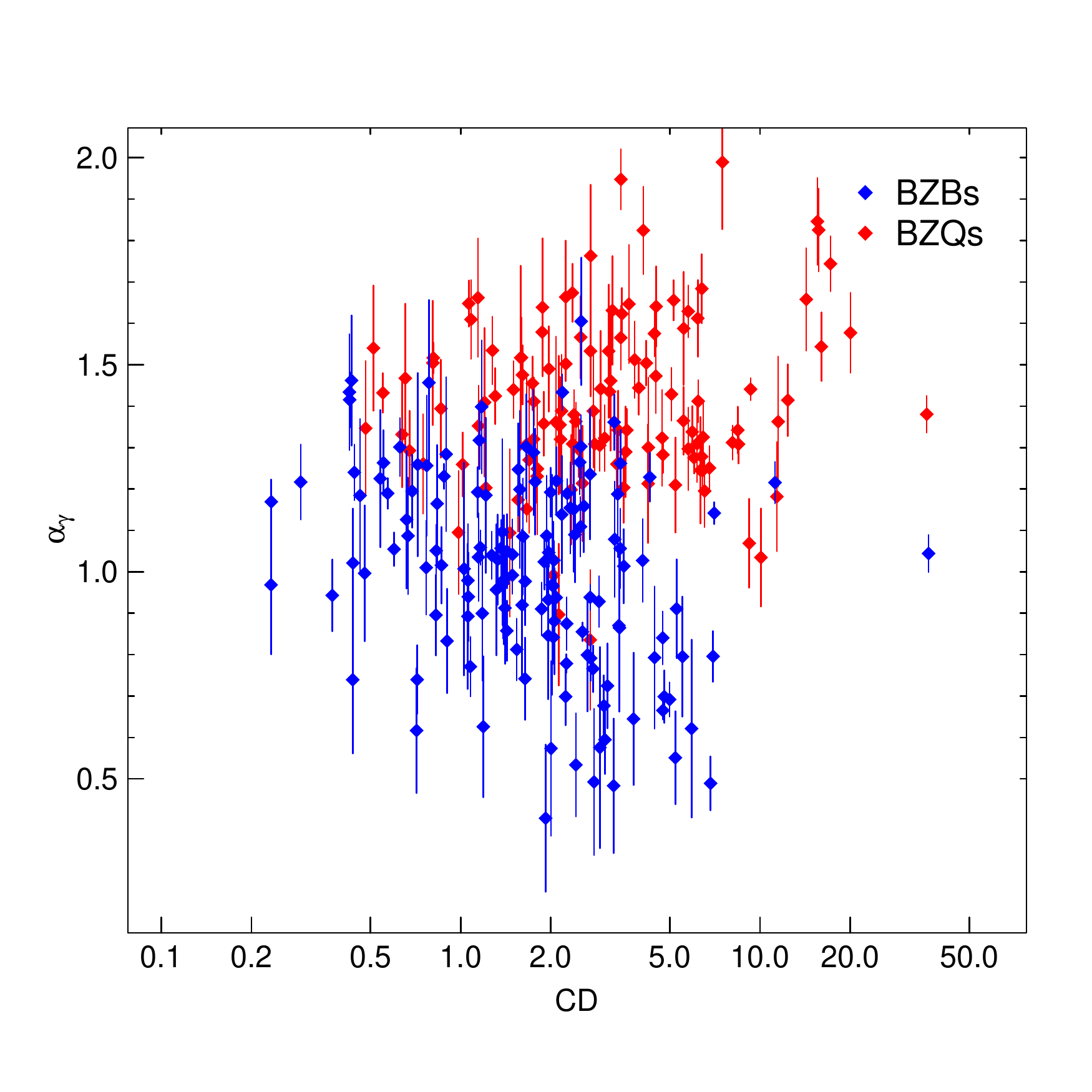}
\caption{Scatterplot of the $\alpha_{\gamma}$ and the CD parameter for the BZBs (blue symbols) 
and the BZQs (red symbols).}
\label{fig:cdvspectralindex_bzbs}
\end{figure}

In Fig.~\ref{fig:histocomptondominancecumulative} we show the two distributions of the CD parameter
for the different classes of BZBs and BZQs. We performed a KS test on the CD distributions of 
BZBs and BZQs and we found that these differ by a 78\% level of significance.
Finally, we report the relation between the CD parameter and the $\alpha_{\gamma}$, 
for both classes of blazars in Fig.~\ref{fig:cdvspectralindex_bzbs}. 

\section{The two families of BL Lac objects}
\label{sec:bllacs}

BL Lacs were originally subclassified in two families on the basis of their radio to X-ray spectral index \citep{padovani95}.
This classification scheme has been recently extended to all types of non-thermal dominated AGNs \citep{abdo10},
on the basis of the position of the peak of the first SED component, generally assumed to be synchrotron emission.
This gives rise to the distinction between the {\it ``Low" - ``Intermediate" - ``High" Synchrotron peaked} non-thermal sources 
(LSPs, ISPs, HSPs), whenever the peak of the synchrotron component lies below 10$^{14}$Hz {\bf($\sim 3 \mu m$)}, between 
10$^{14}$Hz and 10$^{15}$Hz {\bf($\sim 0.3 \mu m$)}, or higher than 10$^{15}$Hz, respectively \citep{abdo10}. 
Even if blazars should most appropriately be classified on the basis of a complete SED, built with simultaneous
data, this is not possible in the majority of the cases, but LSP or HSP BL Lac objects can still be identified by using 
radio-optical-X-ray spectral indices \citep{padovani03,giommi05}. The frequency of the synchrotron peak is 
estimated using the broadband spectral indices between radio and optical wavelengths ($\alpha_{ro}$) and 
the optical and X-ray wavelengths ($\alpha_{ox}$) and extrapolating the spectral shape of BL Lacs to infrared frequency, 
where the peak is expected to be located according to accepted BL Lacs SED model. \wse\ allows to directly
observe the peak or a spectral interval very close to where the synchrotron emission peaks is expected.

\begin{figure}
\includegraphics[height=6.2cm,width=8.5cm,angle=0]{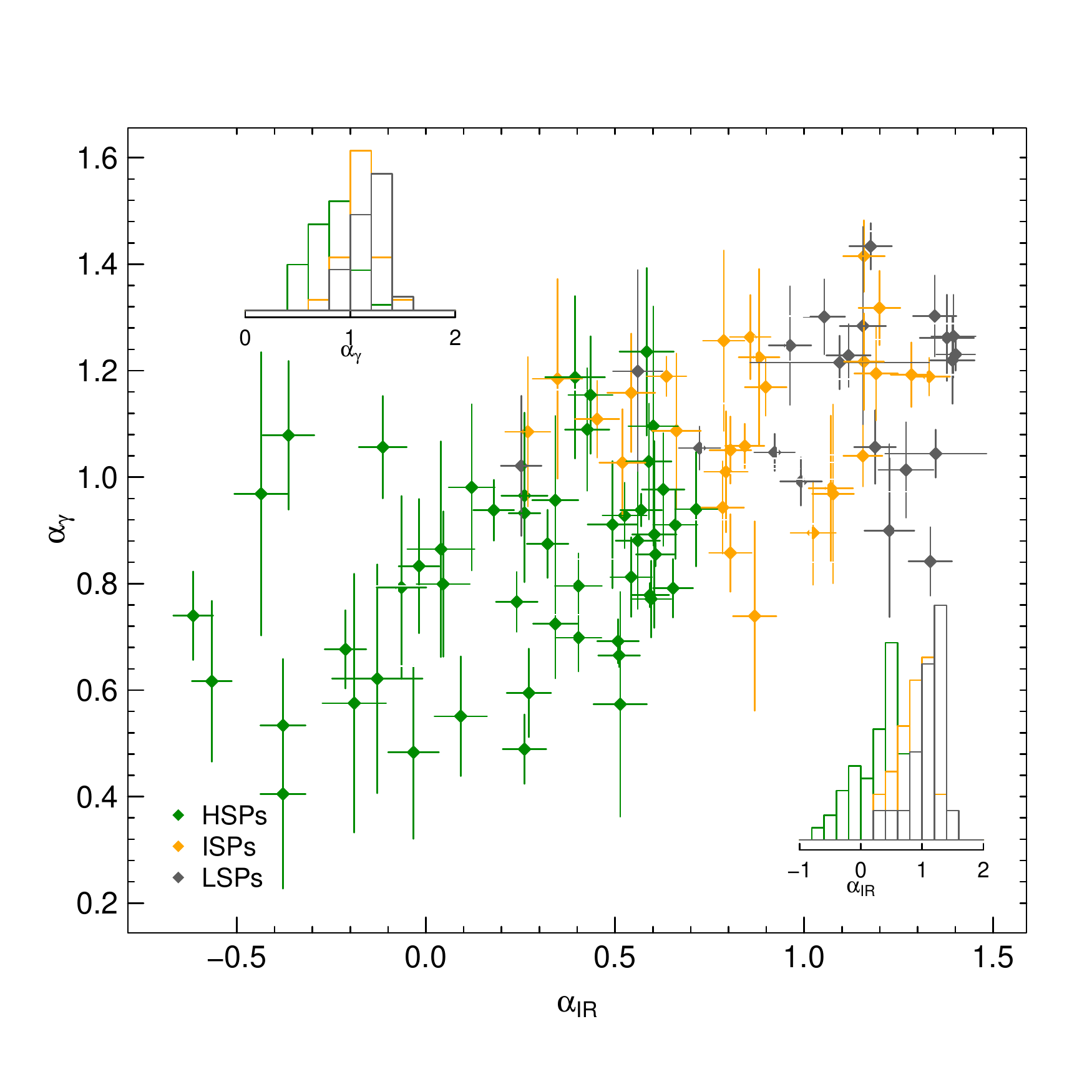}
\caption{Scatterplot of the $\alpha_{IR}-\alpha_{\gamma}$ distribution for BL Lacs.
The three BL Lac subclasses of HSPs (green points), LSP (black points) and ISPs (yellow points) are shown. 
We also report the histogram of the two distributions of $\alpha_{IR}$ and $\alpha_{\gamma}$ 
that clearly show the dichotomy between the three subclasses.}
\label{fig:spectralindexIRspectralindexGAMMA_classed}
\end{figure}

\begin{figure}
\includegraphics[height=6.2cm,width=8.5cm,angle=0]{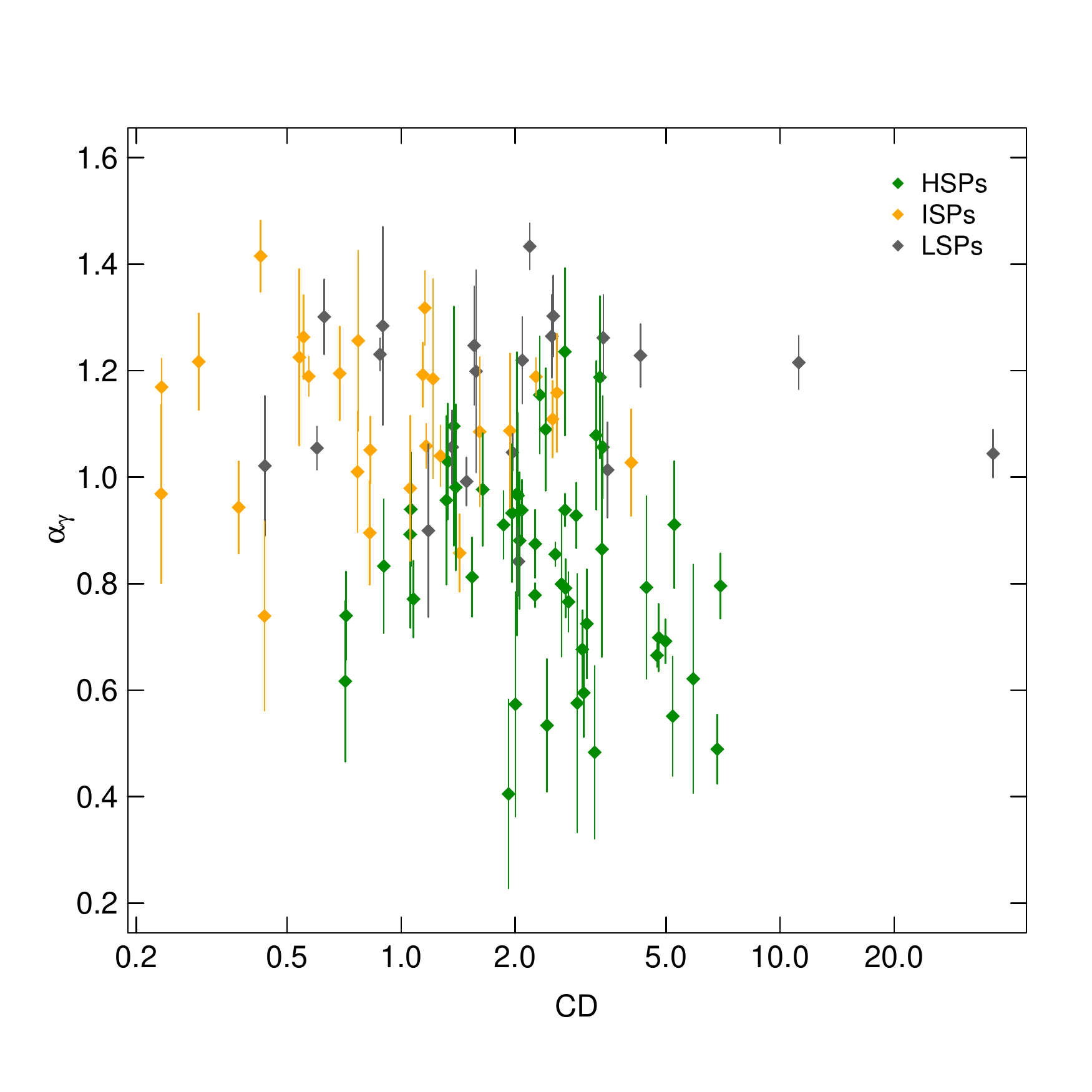}
\caption{Scatterplot of the CD - $\alpha_{\gamma}$ distribution for BL Lacs. The three BL Lac subclasses 
of HSPs (green points), LSP (black points) and ISPs (yellow points) are shown. (see Section~\ref{sec:bllacs} 
for more details).}
\label{fig:CDvspectralindexGAMMA_classed}
\end{figure}

In this paper we adopt the SED classification criterion for the BL Lacs for the \fer\ - \wse\ blazars sample 
distinguishing HSPs, ISPs and LSPs according to the 2LAC sample \citep{ackermann11}. For these three 
families of BL Lacs we studied the relation between the two spectral indices $\alpha_{\gamma}$-$\alpha_{IR}$. 
We notice that, as shown in Fig.~\ref{fig:spectralindexIRspectralindexGAMMA_classed}, there is a clear 
distinction between the two classes in the $\alpha_{\gamma}$, however it is more evident in the 
$\alpha_{IR}$ distribution (a 94\% level of significance has been obtained with a KS test).
A similar distinction between BL Lacs subclasses is visible in the color-magnitude plot produced using the 
[3.4]-[4.6] $\mu$m color and the [3.4] $\mu$m magnitude in Fig.~\ref{fig:color1mag1_classed} in the Appendix
(Section~\ref{sec:appendix}). The three families have clearly separated color distributions, with HSPs, ISPs and 
LSPs with average values of the [3.4]-[4.6] $\mu$m color 0.72, 0.94 and 1.02 respectively.
Finally, we calculate the correlation coefficient  between the variables $\alpha_{\gamma}$ and $\alpha_{IR}$, 
finding $r_s = 0.61$ (p-value negligible), implying that the two variables are correlated within a high level of 
significance. In Fig.~\ref{fig:CDvspectralindexGAMMA_classed}, 
the three different BL Lacs classes in the CD vs $\alpha_{gamma}$ plane are shown. In this case, the CD 
distributions of the HSPs and LSPs are similar (at a 70\% level of confidence, obtained with a KS test), 
while, for the ISPs and the HSPs, the null hypothesis can be rejected to a 87\% level of significance.

\section{Summary and Discussion}
\label{sec:summary}

We have presented the infrared characterization of a sample of blazars detected in the $\gamma$-ray.
In order to perform our selection, we considered all the blazars in the ROMA-BZCAT
catalog \citep{massaro10} that are associated with a $\gamma$-ray source in the 2FGL \citep{abdo11}.
Then, we searched for infrared counterparts in the \wse\
archive adopting the same criteria described in \citep{massaro11b} (see also Section~\ref{sec:sample} 
for more details). The 296 \wse\ counterparts of the ROMA-BZCAT - \fer\ blazars constitute our sample (i.e. the 
\fer\ - \wse\ blazars sample). This more accurate characterization of 
a sample of blazars, as obtained by combining infrared and $\gamma$-ray observations,  
will provide crucial clues for the understanding the unassociated \fer\ objects \citep{massaro11c},
since we expect most of them to be blazars candidates.

We find that the \fer\ - \wse\ blazars cover a very limited region of the [3.4]-[4.6]-[12]-[22] $\mu$m color-color plane,
narrower than the similar locus found for the complete blazars population of the ROMA-BZCAT seen by \wse\ (i.e., the so 
called \strip\ , Massaro et al. 2011b) (see Fig.~\ref{fig:color2color1}).
In particular, we show how the separation between the \fer\ - \wse\ blazars sample and the other extragalactic sources, 
not dominated by synchrotron emission, is evident even with different choices of infrared colors (see Fig.~\ref{fig:color1color3} 
and \ref{fig:color2color3}).
From the three independent infrared colors obtained with \wse\ magnitudes, we have derived the values of the spectral 
indices, finding that the IR spectrum of blazars is clearly consistent with a simple power-law and does not show any 
evidence of deviation from that.

We investigate the properties of the relation between the spectral indices in the $\gamma$-rays and in the infrared.
We found a clear trend between $\alpha_{IR}-\alpha_{\gamma}$ consistent with the expectations of the 
SSC or the EC scenarios. In particular, in the $\alpha_{IR}-\alpha_{\gamma}$ plot the dichotomy between
the two main classes of blazars, BZBs and BZQs is apparent. We also calculate the ratio between the infrared 
flux, integrated over the four \wse\ bands and the total $\gamma$-ray flux as reported in the 2FGL \citep{abdo11}.
This ratio can be used to estimate the Compton Dominance (CD) parameter. We find that the CD distribution for 
the BZB population is more consistent with a synchrotron dominated scenario; the BZQs, as expected, show 
values of CD typically higher than unity, in agreement 
with an inverse Compton framework and with the widely accepted EC emission model.

We also considered the BZB subclasses (i.e., HSPs, ISPs, LSPs) as defined in \citep{abdo10}
and we investigated their IR-to-$\gamma$-ray properties. We find a strong correlation between 
the spectral indices and the classification in BZB families; in particular, the HSPs class appears 
to be very different from the LSPs in the $\alpha_{IR}$ distribution. As already shown in 
\citep{massaro11b}, blazars can be separated in the \wse\ IR colors from other sources not 
dominated by synchrotron emission, however this distinction appears to be more evident when 
considering those selected on the basis of their $\gamma$-ray properties.
For this reason, while we are aware that the results discuss in this paper cannot be generalized, in principle, 
to all $\gamma$ emitting blazars because of the inhomogeneity of the parent catalog, we deem the peculiar IR features 
of this large sample of confirmed blazars worth to be investigated and interesting \emph{per se}. Moreover, since 
blazars constitute the most detected extragalactic 
sources at $\gamma$-ray energies, unidentified \fer\ AGN candidates or \fer\ unassociated objects \citep{abdo11} 
are likely to be unknown blazar. For this reason, we are investigating the possibility of employing this new infrared 
correlations as diagnostic tools to associate otherwise unassociated \fer\ sources with blazars, and so better categorize this
interesting class of extragalactic sources \citep{massaro11c}.

\acknowledgements
R. D'Abrusco gratefully acknowledges the financial support of the US Virtual Astronomical Observatory, 
which is sponsored by the National Science Foundation and the National Aeronautics and Space Administration.
F. Massaro is grateful to D. Harris for their constructive discussions that improved the presentation of the paper.
The work at SAO is partially supported by the NASA grant NNX10AD50G and NNX10AD68G. F. Massaro also 
acknowledges the Fondazione Angelo Della Riccia for the grant awarded him to support his research at SAO 
during 2011 and the Foundation BLANCEFLOR Boncompagni-Ludovisi, n\'ee Bildt for the grant awarded him 
in 2010 to support his research. M. Ajello and F. Massaro acknowledge support from NASA grant NNH09ZDA001N.
TOPCAT\footnote{\underline{http://www.star.bris.ac.uk/$\sim$mbt/topcat/}} \citep{taylor05} was used extensively 
in this work for the preparation and manipulation of the tabular data.
Part of this work is based on archival data, software or on-line services provided by the ASI Science Data Center.
This publication makes use of data products from the Wide-field Infrared Survey Explorer, which is a joint project of the 
University of California, Los Angeles, and the Jet Propulsion Laboratory/California Institute of Technology, funded by the 
National Aeronautics and Space Administration.

\appendix

\section{Additional plots}
\label{sec:appendix}

\begin{figure}
\begin{center}
\includegraphics[height=10cm,width=10cm,angle=0]{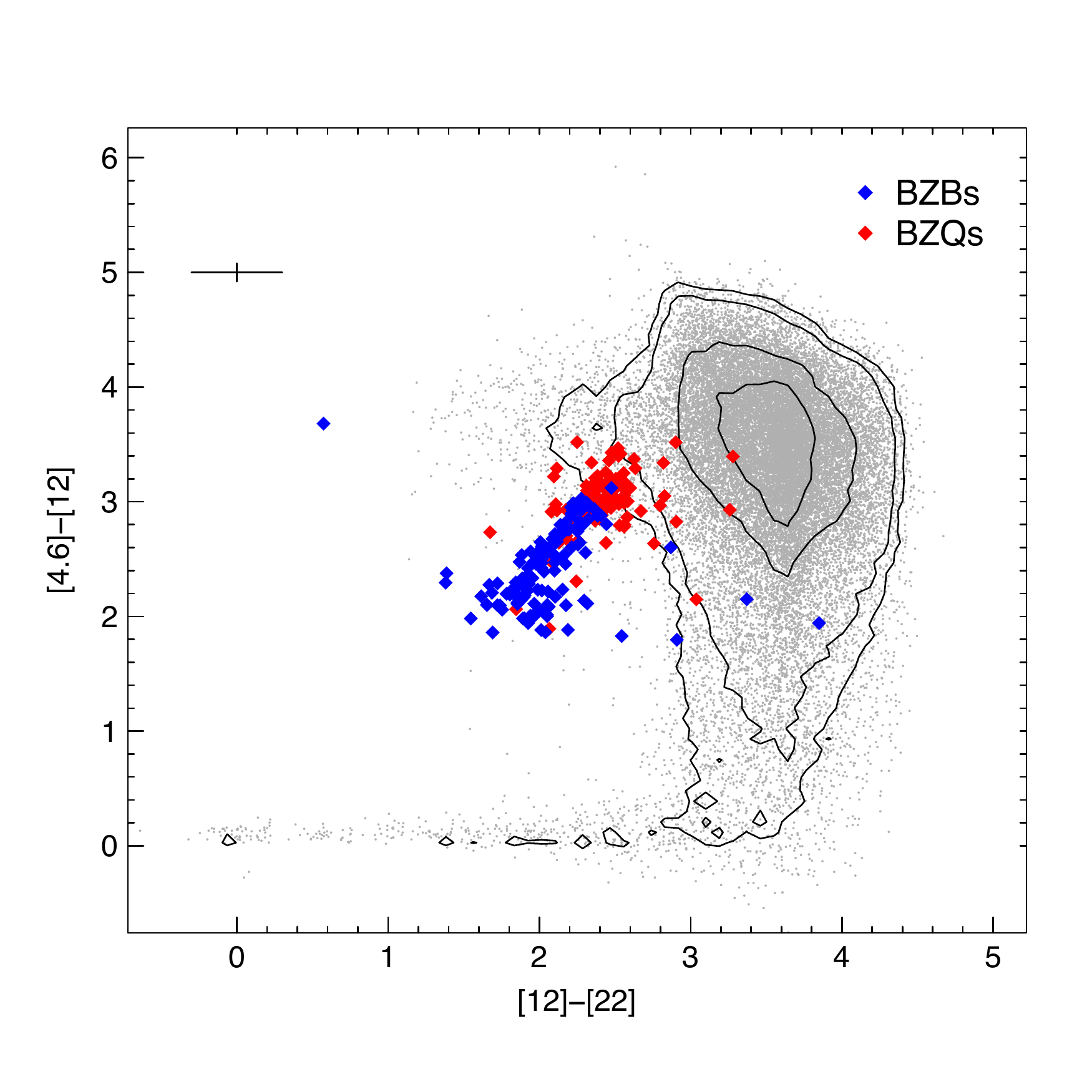}
\caption{The [4.6]-[12]-[22] $\mu$m color-color diagram of \fer\ - \wse\ blazars sample sources.
The two blazars classes of BZBs (blue) and BZQs (red) are shown. The background grey dots correspond 
to 453420 \wse\ sources detected in a region of 56 deg$^2$ at high Galactic latitude. The isodensity 
curves for the \wse\ sources, corresponding to 50, 100, 500, 2000 sources per unit area in the color-color 
plane respectively, are shown (see Section~\ref{sec:2FBS}). The black cross shown in the right bottom 
represents the typical error on the infrared colors.}
\label{fig:color2color3}
\end{center}
\end{figure}

\begin{figure}
\begin{center}
\includegraphics[height=10cm,width=10cm,angle=0]{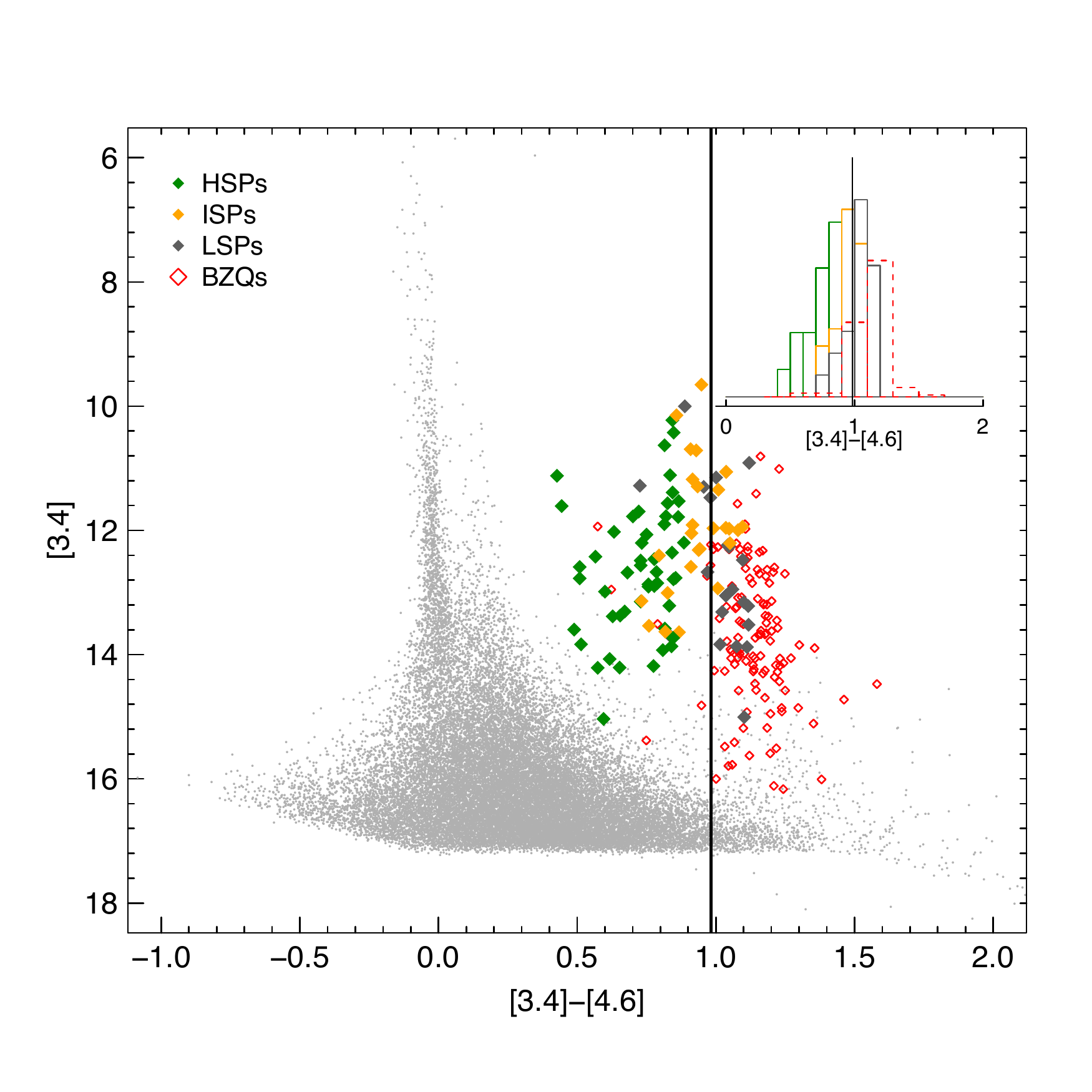}
\caption{The [3.4]-[4.6]  vs [3.4] color-magnitude diagram of the sources in the \fer\ - \wse\ blazars sample. 
The BZQs (red symbols) and the HSPs, ISPs and LSPs BL Lacs classes (see 
section~\ref{sec:bllacs}) are shown (green, orange and gray symbols respectively). The black vertical 
line corresponds to the color value associated to a power law with spectral index $\alpha_{\nu}=-1$.}
\label{fig:color1mag1_classed}
\end{center}
\end{figure}

\begin{figure}
\begin{center}
\begin{tabular}{cc}
\includegraphics[width=0.4\linewidth]{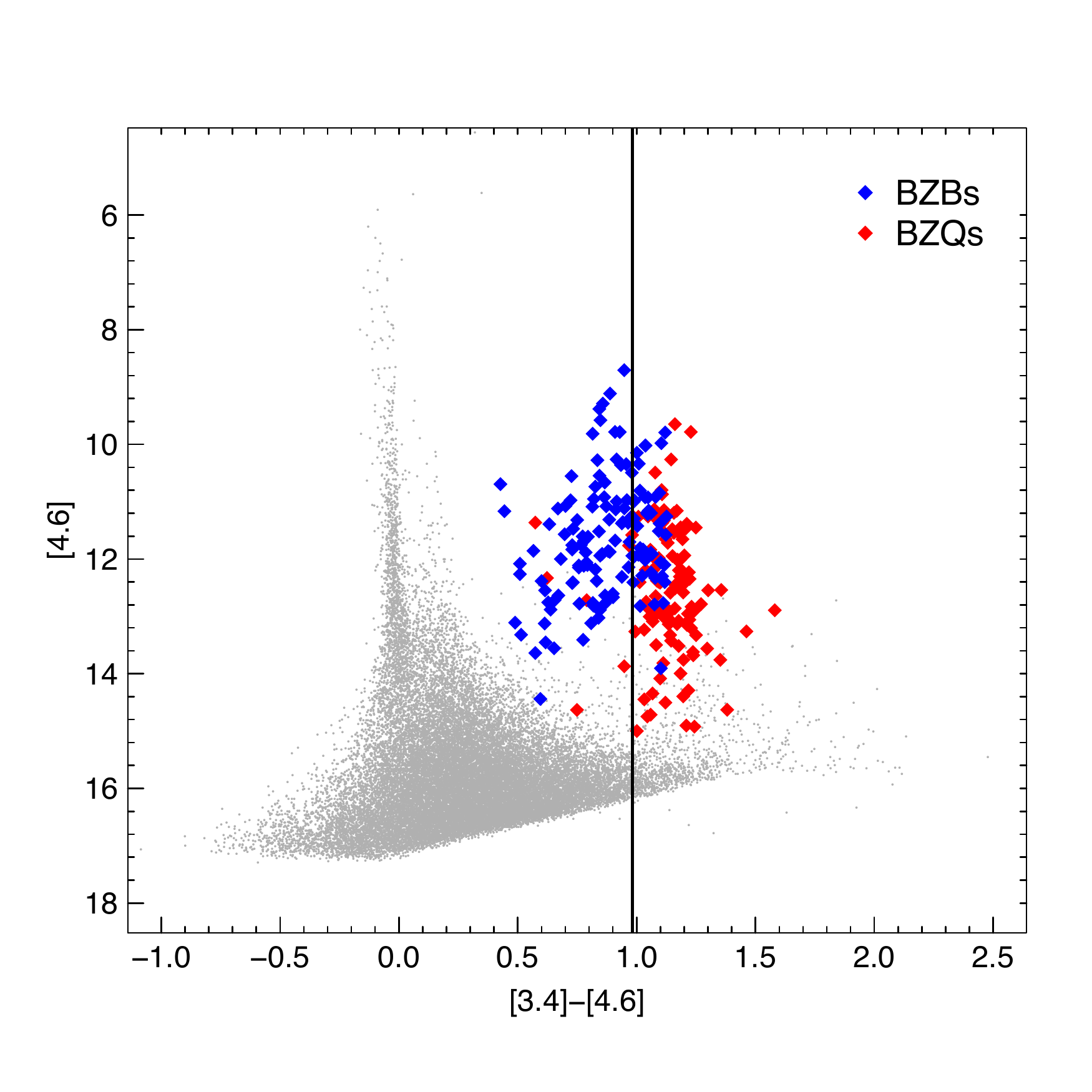} &
\includegraphics[width=0.4\linewidth]{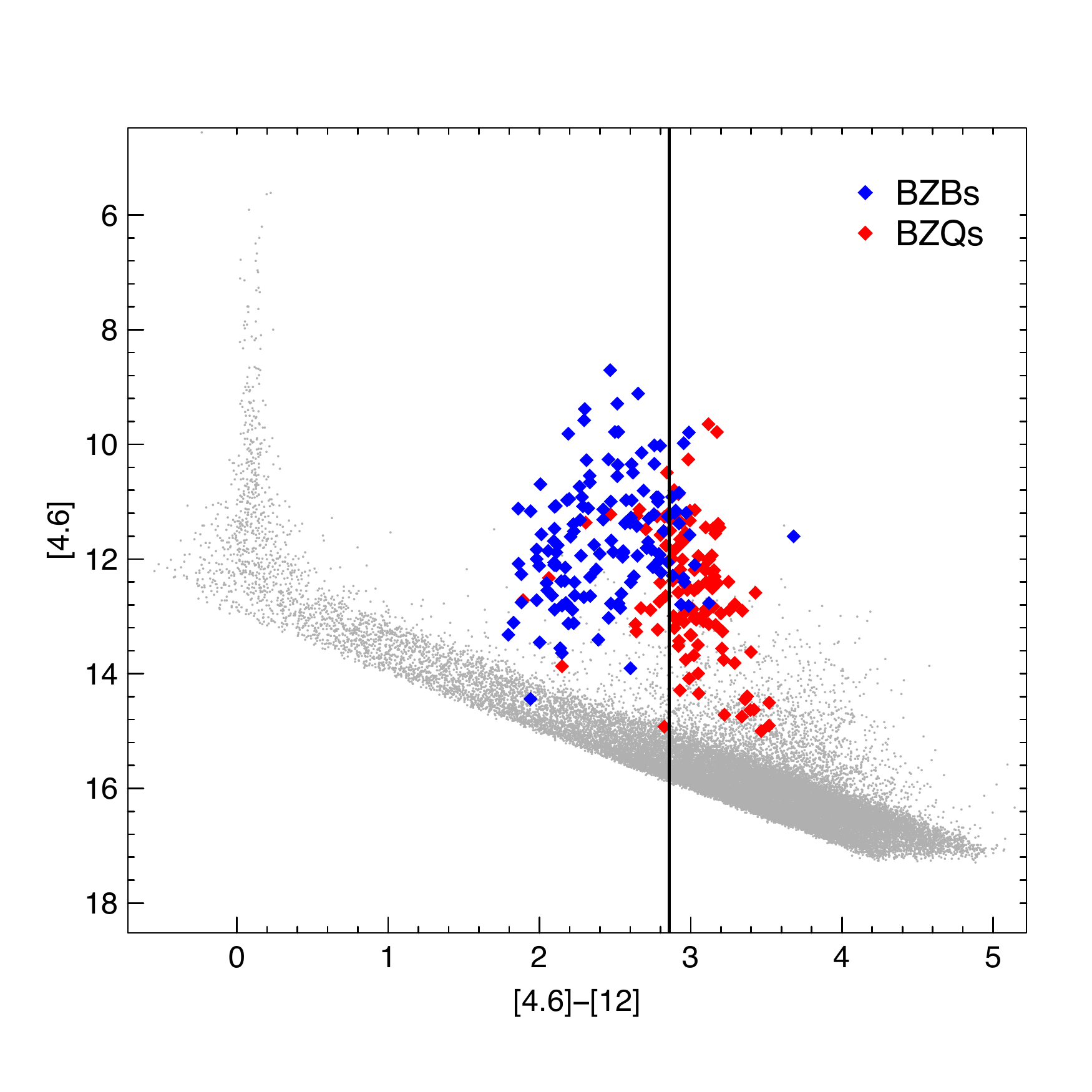} \\
\includegraphics[width=0.4\linewidth]{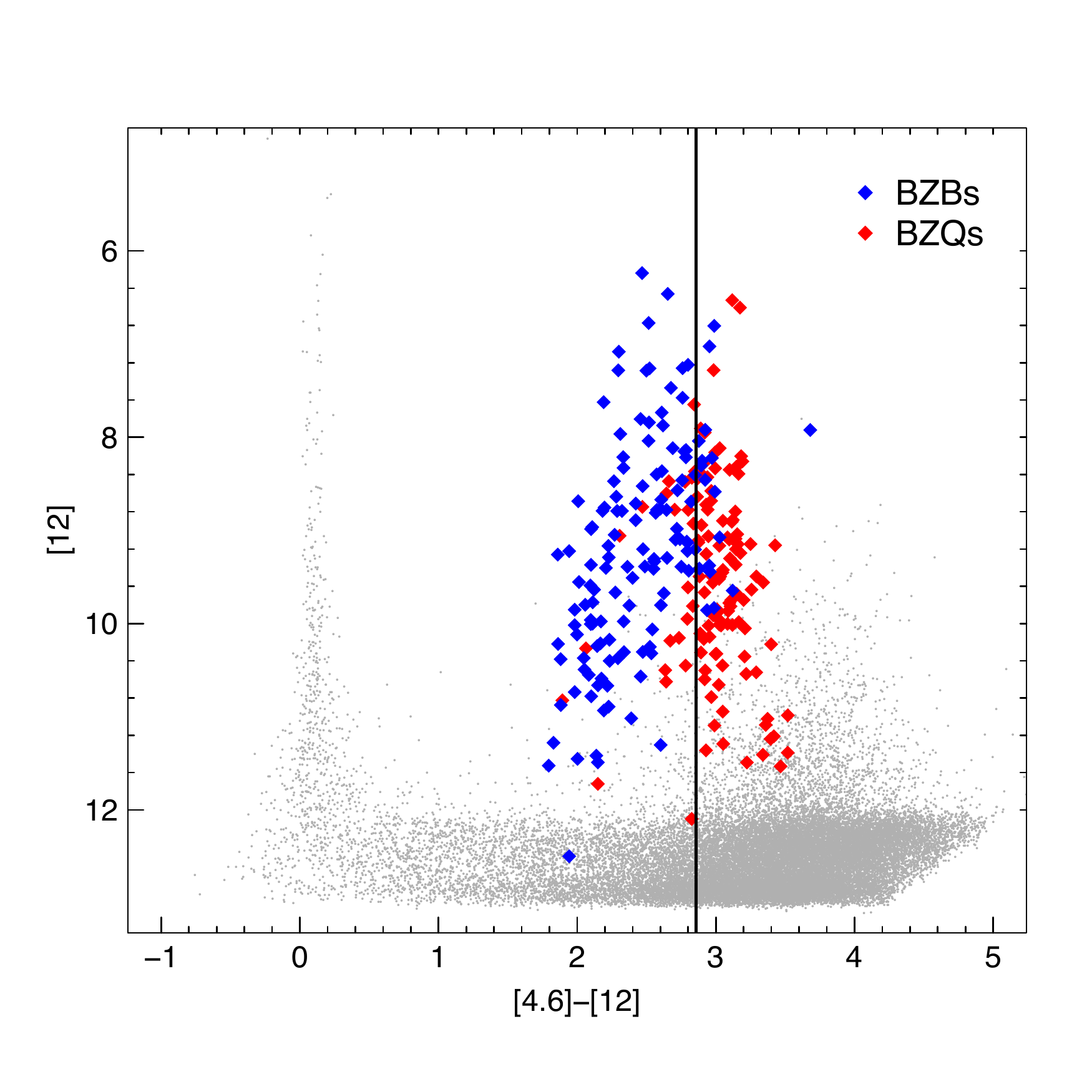} &
\includegraphics[width=0.4\linewidth]{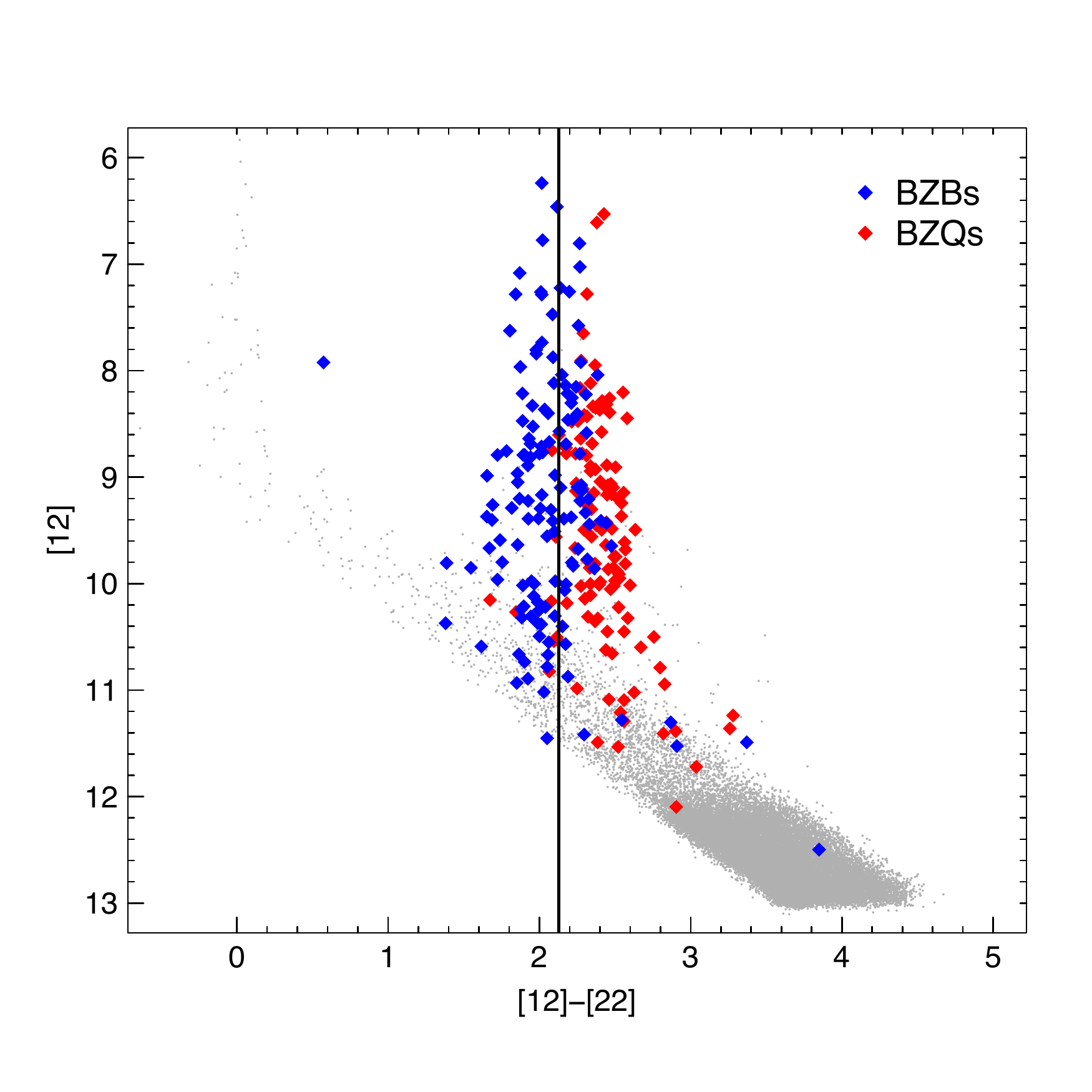} \\
\includegraphics[width=0.4\linewidth]{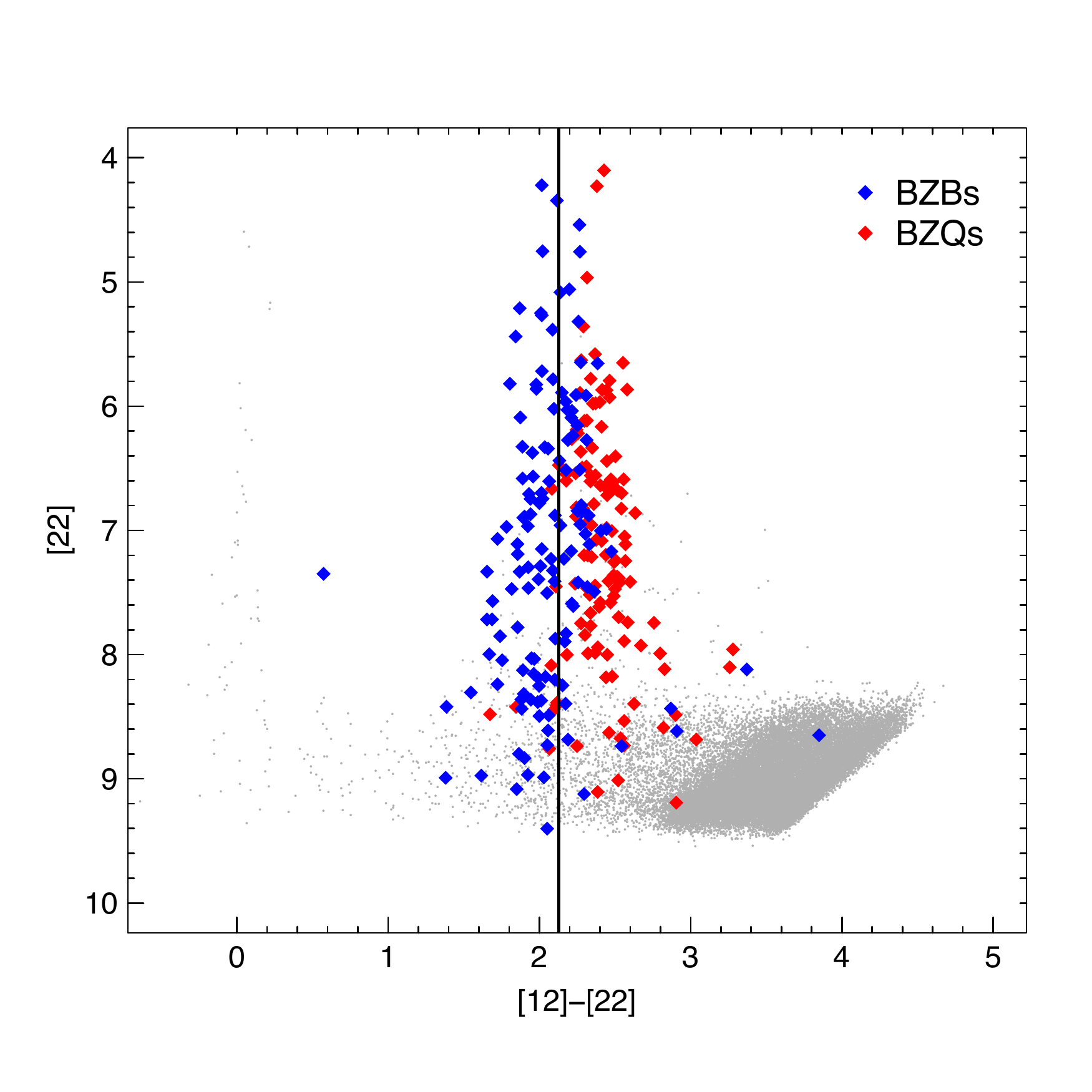} &
\end{tabular}
\end{center}
\caption{Color-magnitude diagrams of the \fer\ - \wse\ sample sources. From top left, [3.4]-[4.6] vs [4.6] plot, 
[4.6]-[12] vs [4.6] plot, [4.6]-[12] vs [12] plot, [12]-[22] vs [12] plot and [12]-[22] vs [22] plot. 
The BZQs (red symbols) and BZBs (blue symbols) are shown in all plots. The black vertical 
lines correspond to the color value associated to a power law with spectral index $\alpha_{\nu}=-1$.}
\label{fig:colorsmags}
\end{figure}

In this appendix, we present additional plots for the \fer\ - \wse\ blazars sample, namely the [4.6]-[12]-[33] color-color plot
(Fig.~\ref{fig:color2color3}) for BZBs and BZQs, the color-magnitude plot built using [3.4]-[4.6] color and [3.4] magnitude 
with the different classes of BL Lacs (Fig.~\ref{fig:color1mag1_classed}), and several color-magnitude plots with BZBs 
and BZQs (Fig.~\ref{fig:colorsmags}) obtained using \wse\ photometric data.   

\clearpage

\section{Table}
\label{sec:appendix2}

\begin{center}
\begin{longtable}{ccrrrrrr}
\caption{Table of parameters of the \fer\ - \wse\ blazars sample. The first column contains the name of the blazer in the 
ROMA-BZCat catalog, the second column the name of the associated source from the 2FGL catalog, the columns from 
the third to the sixth contain the values of the magnitude in the four \wse\ filters ([3.4], [4.6], [12] and [22] $\mu$m respectively), 
the seventh column contains the $\gamma$ spectral index of the sources from the 2FGL catalog and the last column
the IR spectral index $\alpha_{12}$ evaluated from \wse\ magnitudes [3.4] and [4.6].) Errors are reported in 
parenthesis.}\label{tab:blazars}\\
\tableline\tableline
  ROMA-BZCat name & 2FGL name & m$_1$ & m$_2$ & m$_3$ & m$_4$ & $\Gamma_{\mathrm{\gamma}}$ & $\alpha_{\mathrm{IR}}$\\ 
\tableline  
\endfirsthead  

%\tableline\tableline
%  ROMA-BZCat name & 2FGL name & m$_1$ & m$_2$ & m$_3$ & m$_4$ & $\alpha_{\mathrm{\gamma}}$ & $\alpha_{\mathrm{IR}}$\\ 
%\tableline  
%\endhead
\tableline
\endhead

\tableline
\endlastfoot
  BZBJ0035+5950 & 1ES 0033+595 & 12.26(0.03) & 11.57(0.02) & 9.56(0.05) & 7.5(0.15) & 1.87(0.07) & 0.17(0.06)\\
  BZBJ0123+3420 & 1ES 0120+340 & 13.92(0.02) & 13.31(0.03) & 11.5(0.21) & 8.29() & 1.53(0.21) & -0.09(0.06)\\
  BZBJ0136+3905 & B3 0133+388 & 11.9(0.02) & 11.09(0.02) & 8.99(0.03) & 7.33(0.09) & 1.69(0.04) & 0.51(0.06)\\
  BZBJ0154+0823 & GB6 J0154+0823 & 11.91(0.02) & 11.0(0.02) & 8.52(0.03) & 6.57(0.07) & 1.86(0.07) & 0.81(0.06)\\
  BZBJ0154+4433 & GB6 J0154+4433 & 13.25(0.03) & 12.31(0.03) & 9.98(0.05) & 8.03(0.2) & 2.01(0.26) & 0.87(0.06)\\
  BZBJ0159+1047 & RX J0159.5+1047 & 12.85(0.02) & 12.06(0.02) & 9.96(0.06) & 8.24(0.3) & 2.15(0.11) & 0.44(0.06)\\
  BZBJ0203+7232 & S5 0159+723 & 11.28(0.02) & 10.55(0.02) & 8.04(0.02) & 5.89(0.04) & 2.02(0.13) & 0.25(0.05)\\
  BZBJ0212+2244 & MG3 J021252+2246 & 12.36(0.02) & 11.52(0.02) & 9.29(0.05) & 7.47(0.21) & 2.03(0.11) & 0.59(0.06)\\
  BZBJ0217+0837 & ZS 0214+083 & 10.69(0.02) & 9.78(0.02) & 7.26(0.02) & 5.25(0.03) & 1.94(0.09) & 0.78(0.06)\\
  BZBJ0222+4302 & 3C 66A & 9.55(0.02) & 8.7(0.02) & 6.42(0.02) & 4.54(0.03) & 1.85(0.02) & 0.6(0.05)\\
  BZBJ0238+1636 & AO 0235+164 & 11.78(0.02) & 10.66(0.02) & 7.7(0.02) & 5.39(0.03) & 2.02(0.02) & 1.39(0.05)\\
  BZBJ0243+7120 & S5 0238+711 & 12.95(0.03) & 11.89(0.02) & 9.33(0.03) & 7.03(0.09) & 1.9(0.16) & 1.23(0.07)\\
  BZBJ0258+2030 & MG3 J025805+2029 & 12.84(0.03) & 11.82(0.02) & 9.48(0.05) & 7.52(0.22) & 2.19(0.17) & 1.1(0.06)\\
  BZBJ0303-2407 & PKS 0301-243 & 11.11(0.02) & 10.28(0.02) & 7.96(0.02) & 6.09(0.05) & 1.94(0.03) & 0.57(0.06)\\
  BZBJ0303+4716 & 4C +47.08 & 11.08(0.02) & 9.98(0.02) & 7.02(0.02) & 4.76(0.02) & 2.24(0.07) & 1.35(0.05)\\
  BZBJ0304-2832 & RBS 0385 & 15.04(0.04) & 14.44(0.06) & 12.5() & 8.65() & 1.62(0.21) & -0.13(0.12)\\
  BZBJ0316-2607 & RBS 0405 & 12.99(0.03) & 12.24(0.02) & 10.06(0.05) & 8.45(0.24) & 1.87(0.14) & 0.33(0.06)\\
  BZBJ0316+0904 & GB6 J0316+0904 & 11.56(0.02) & 10.74(0.02) & 8.47(0.03) & 6.58(0.07) & 1.81(0.07) & 0.54(0.06)\\
  BZBJ0319+1845 & RBS 0413 & 13.31(0.03) & 12.64(0.03) & 10.55(0.09) & 8.49() & 1.55(0.11) & 0.09(0.07)\\
  BZBJ0321+2326 & MG3 J032201+2336 & 12.67(0.03) & 11.88(0.02) & 9.77(0.05) & 7.46(0.13) & 2.09(0.12) & 0.43(0.06)\\
  BZBJ0325-1646 & RBS 0421 & 13.15(0.02) & 12.42(0.03) & 10.37(0.06) & 8.38(0.27) & 1.97(0.16) & 0.26(0.06)\\
  BZBJ0326+0225 & 1H 0323+022 & 12.98(0.03) & 12.38(0.03) & 10.24(0.07) & 8.36(0.28) & 2.06(0.1) & -0.11(0.06)\\
  BZBJ0334-4008 & PKS 0332-403 & 11.94(0.02) & 10.84(0.02) & 7.92(0.02) & 5.65(0.03) & 2.19(0.04) & 1.33(0.06)\\
  BZBJ0334-3725 & PMN J0334-3725 & 11.47(0.02) & 10.49(0.02) & 7.87(0.02) & 5.78(0.04) & 1.99(0.05) & 0.99(0.06)\\
  BZBJ0340-2119 & PKS 0338-214 & 11.99(0.02) & 11.0(0.02) & 8.22(0.02) & 6.03(0.04) & 2.43(0.14) & 1.02(0.06)\\
  BZBJ0357-4955 & PKS 0355-500 & 12.32(0.02) & 11.38(0.02) & 8.81(0.02) & 6.87(0.06) & 1.74(0.18) & 0.87(0.06)\\
  BZBJ0424+0036 & PKS 0422+00 & 11.15(0.02) & 10.15(0.02) & 7.47(0.02) & 5.38(0.03) & 2.3(0.07) & 1.05(0.06)\\
  BZBJ0428-3756 & PKS 0426-380 & 10.84(0.02) & 9.81(0.02) & 7.11(0.02) & 4.9(0.03) & 1.95(0.02) & 1.13(0.05)\\
  BZBJ0430-2507 & PMN J0430-2507 & 13.69(0.03) & 12.85(0.03) & 10.32(0.06) & 8.44(0.27) & 2.2(0.19) & 0.56(0.07)\\
  BZBJ0433+2905 & MG2 J043337+2905 & 15.01(0.06) & 13.9(0.06) & 11.3(0.18) & 8.43() & 2.04(0.05) & 1.35(0.13)\\
  BZBJ0434-2015 & TXS 0431-203 & 13.39(0.03) & 12.41(0.03) & 9.8(0.04) & 7.59(0.14) & 2.22(0.13) & 1.01(0.06)\\
  BZBJ0448-1632 & RBS 0589 & 13.93(0.03) & 13.12(0.03) & 10.89(0.08) & 8.97(0.33) & 1.91(0.12) & 0.49(0.07)\\
  BZBJ0449-4350 & PKS 0447-439 & 10.43(0.02) & 9.58(0.02) & 7.28(0.02) & 5.44(0.03) & 1.86(0.02) & 0.61(0.05)\\
  BZBJ0507+6737 & 1ES 0502+675 & 12.56(0.03) & 11.83(0.02) & 9.85(0.04) & 8.3(0.23) & 1.49(0.07) & 0.26(0.06)\\
  BZBJ0509+0541 & TXS 0506+056 & 10.71(0.02) & 9.78(0.02) & 7.28(0.02) & 5.27(0.03) & 2.06(0.04) & 0.84(0.05)\\
  BZBJ0536-3343 & 1RXS J053629.4-334302 & 12.98(0.02) & 12.28(0.02) & 10.22(0.05) & 7.95(0.15) & 2.39(0.06) & 0.19(0.06)\\
  BZBJ0538-4405 & PKS 0537-441 & 9.09(0.02) & 8.0(0.02) & 5.21(0.02) & 3.04(0.02) & 2.01(0.02) & 1.3(0.05)\\
  BZBJ0558-7459 & PKS 0600-749 & 12.25(0.02) & 11.28(0.02) & 8.67(0.02) & 6.6(0.05) & 2.09(0.14) & 0.98(0.06)\\
  BZBJ0607+4739 & TXS 0603+476 & 11.18(0.02) & 10.26(0.02) & 7.8(0.02) & 5.83(0.04) & 2.05(0.06) & 0.81(0.06)\\
  BZBJ0612+4122 & B3 0609+413 & 11.82(0.03) & 10.81(0.02) & 8.12(0.02) & 6.02(0.04) & 2.03(0.05) & 1.09(0.06)\\
  BZBJ0617+5701 & 87GB 061258.1+570222 & 11.97(0.03) & 10.98(0.02) & 8.36(0.02) & 6.33(0.05) & 1.9(0.1) & 1.02(0.06)\\
  BZBJ0625+4440 & GB6 J0625+4440 & 12.92(0.03) & 11.94(0.02) & 9.3(0.03) & 7.29(0.1) & 1.91(0.14) & 1.0(0.06)\\
  BZBJ0629-1959 & PKS 0627-199 & 12.38(0.03) & 11.26(0.02) & 8.41(0.02) & 6.16(0.04) & 2.19(0.06) & 1.42(0.06)\\
  BZBJ0650+2503 & 1ES 0647+250 & 12.2(0.03) & 11.47(0.02) & 9.37(0.04) & 7.72(0.17) & 1.59(0.08) & 0.27(0.06)\\
  BZBJ0710+5908 & 1H 0658+595 & 12.59(0.02) & 12.08(0.02) & 10.22(0.06) & 8.18(0.19) & 1.53(0.12) & -0.38(0.06)\\
  BZBJ0712+5033 & GB6 J0712+5033 & 12.27(0.02) & 11.22(0.02) & 8.46(0.02) & 6.27(0.06) & 2.06(0.07) & 1.19(0.06)\\
  BZBJ0721+7120 & S5 0716+71 & 9.1(0.02) & 8.13(0.02) & 5.44(0.02) & 3.4(0.02) & 2.01(0.02) & 0.98(0.05)\\
  BZBJ0730+3307 & 1RXS J073026.0+330727 & 12.79(0.03) & 11.94(0.02) & 9.67(0.05) & 8.0(0.2) & 1.89(0.18) & 0.6(0.06)\\
  BZBJ0738+1742 & PKS 0735+17 & 11.3(0.02) & 10.34(0.02) & 7.74(0.02) & 5.72(0.04) & 2.05(0.03) & 0.92(0.05)\\
  BZBJ0744+7433 & MS 0737.9+7441 & 13.37(0.03) & 12.72(0.03) & 10.74(0.09) & 8.83() & 1.8(0.14) & 0.05(0.07)\\
  BZBJ0753+5352 & 4C +54.15 & 13.87(0.03) & 12.79(0.03) & 9.86(0.05) & 7.49(0.13) & 2.01(0.09) & 1.27(0.07)\\
  BZBJ0754+4823 & GB1 0751+485 & 11.97(0.03) & 10.92(0.02) & 8.15(0.02) & 5.91(0.04) & 2.19(0.09) & 1.19(0.06)\\
  BZBJ0757+0956 & PKS 0754+100 & 12.0(0.02) & 10.92(0.02) & 8.04(0.02) & 5.66(0.04) & 2.19(0.06) & 1.28(0.06)\\
  BZBJ0805+7534 & RX J0805.4+7534 & 12.42(0.02) & 11.86(0.02) & 9.8(0.04) & 8.04(0.16) & 1.68(0.07) & -0.21(0.06)\\
  BZBJ0809+5218 & 1ES 0806+524 & 11.78(0.02) & 11.07(0.02) & 8.96(0.03) & 7.11(0.11) & 1.94(0.06) & 0.18(0.05)\\
  BZBJ0811+0146 & OJ 014 & 13.88(0.03) & 12.77(0.03) & 9.65(0.05) & 7.17(0.1) & 2.26(0.08) & 1.38(0.07)\\
  BZBJ0814+6431 & GB6 J0814+6431 & 11.29(0.02) & 10.36(0.02) & 7.84(0.02) & 5.86(0.04) & 2.26(0.08) & 0.86(0.05)\\
  BZBJ0816-1311 & PMN J0816-1311 & 12.9(0.03) & 12.12(0.02) & 10.12(0.05) & 8.15(0.21) & 1.8(0.06) & 0.4(0.06)\\
  BZBJ0816+5739 & SBS 0812+578 & 12.76(0.02) & 11.91(0.02) & 9.51(0.04) & 7.41(0.1) & 1.98(0.11) & 0.63(0.06)\\
  BZBJ0817+3243 & RX J0817.9+3243 & 14.18(0.03) & 13.41(0.04) & 11.02(0.12) & 8.99(0.44) & 2.19(0.15) & 0.39(0.08)\\
  BZBJ0818+4222 & S4 0814+42 & 12.7(0.03) & 11.58(0.02) & 8.59(0.03) & 6.27(0.05) & 2.14(0.03) & 1.41(0.06)\\
  BZBJ0817-0933 & TXS 0815-094 & 12.43(0.03) & 11.42(0.02) & 8.78(0.03) & 6.51(0.06) & 2.04(0.09) & 1.06(0.06)\\
  BZBJ0819+2747 & 5C 07.119 & 13.3(0.03) & 12.24(0.02) & 9.43(0.04) & 6.99(0.08) & 2.26(0.22) & 1.23(0.06)\\
  BZBJ0819-0756 & RX J0819.2-0756 & 14.21(0.03) & 13.64(0.04) & 11.49(0.16) & 8.12(0.26) & 1.58(0.24) & -0.19(0.08)\\
  BZBJ0834+4403 & B3 0831+442 & 12.85(0.03) & 11.91(0.02) & 9.31(0.04) & 7.2(0.09) & 2.04(0.19) & 0.87(0.06)\\
  BZBJ0839+1802 & TXS 0836+182 & 12.76(0.03) & 11.88(0.02) & 9.39(0.04) & 7.39(0.12) & 2.46(0.2) & 0.72(0.06)\\
  BZBJ0839+3540 & FIRST J083943.3+354001 & 13.54(0.03) & 12.64(0.03) & 10.31(0.08) & 8.36(0.33) & 1.92(0.18) & 0.73(0.07)\\
  BZBJ0844+5312 & BZB J0844+5312 & 13.64(0.03) & 12.77(0.03) & 10.25(0.06) & 8.25(0.3) & 2.09(0.15) & 0.66(0.06)\\
  BZBJ0847+1133 & RX J0847.1+1133 & 13.38(0.03) & 12.76(0.03) & 10.87(0.12) & 8.68(0.42) & 1.48(0.16) & -0.03(0.07)\\
  BZBJ0856+2057 & PMN J0856-1105 & 13.5(0.03) & 12.6(0.03) & 10.06(0.08) & 7.9() & 2.14(0.14) & 0.76(0.07)\\
  BZBJ0902+2050 & NVSS J090226+205045 & 12.05(0.03) & 11.13(0.02) & 8.71(0.03) & 6.7(0.07) & 2.01(0.11) & 0.79(0.06)\\
  BZBJ0910+3329 & Ton 1015 & 12.2(0.02) & 11.31(0.02) & 8.89(0.03) & 6.97(0.08) & 1.94(0.11) & 0.71(0.06)\\
  BZBJ0915+2933 & B2 0912+29 & 12.07(0.02) & 11.32(0.02) & 9.05(0.03) & 7.19(0.1) & 1.87(0.06) & 0.32(0.06)\\
  BZBJ0929+5013 & GB6 J0929+5013 & 12.93(0.03) & 11.93(0.02) & 9.13(0.03) & 6.85(0.06) & 1.98(0.14) & 1.07(0.06)\\
  BZBJ0929+8612 & S5 0916+864 & 12.86(0.02) & 11.84(0.02) & 9.09(0.03) & 6.84(0.06) & 2.05(0.09) & 1.13(0.06)\\
  BZBJ0940+6148 & RX J0940.3+6148 & 13.83(0.03) & 13.32(0.03) & 11.52(0.14) & 8.62() & 2.08(0.14) & -0.36(0.07)\\
  BZBJ0945+5757 & GB6 J0945+5757 & 12.52(0.02) & 11.75(0.02) & 9.39(0.03) & 7.46(0.08) & 2.16(0.14) & 0.37(0.06)\\
  BZBJ0958+6533 & S4 0954+65 & 11.06(0.02) & 10.02(0.02) & 7.22(0.02) & 5.08(0.03) & 2.42(0.07) & 1.16(0.06)\\
  BZBJ1018+5911 & TXS 1015+594 & 13.54(0.03) & 12.78(0.03) & 10.3(0.06) & 8.2(0.21) & 2.18(0.19) & 0.35(0.07)\\
  BZBJ1019+6320 & GB6 J1019+6319 & 12.82(0.02) & 11.81(0.02) & 9.1(0.03) & 6.96(0.06) & 2.18(0.19) & 1.09(0.06)\\
  BZBJ1058-8003 & PKS 1057-79 & 11.52(0.02) & 10.45(0.02) & 7.6(0.02) & 5.4(0.03) & 2.05(0.09) & 1.27(0.05)\\
  BZBJ1110+7133 & 87GB 110723.4+715023 & 13.73(0.03) & 12.89(0.03) & 10.67(0.07) & 8.61(0.26) & 2.1(0.22) & 0.6(0.07)\\
  BZBJ1136+7009 & Mkn 180 & 11.12(0.02) & 10.69(0.02) & 8.69(0.02) & 6.75(0.06) & 1.74(0.08) & -0.62(0.05)\\
  BZBJ1223+8040 & S5 1221+80 & 13.52(0.03) & 12.4(0.02) & 9.44(0.03) & 7.11(0.07) & 2.26(0.08) & 1.4(0.06)\\
  BZBJ1312-2156 & PKS 1309-216 & 12.06(0.03) & 11.12(0.02) & 8.79(0.03) & 6.9(0.1) & 2.02(0.07) & 0.9(0.06)\\
  BZBJ1352-4412 & PKS 1349-439 & 12.6(0.03) & 11.51(0.02) & 8.69(0.03) & 6.52(0.07) & 2.13(0.17) & 1.32(0.06)\\
  BZBJ1357+0128 & RX J1357.6+0128 & 13.68(0.03) & 12.85(0.03) & 10.45(0.08) & 8.46(0.37) & 2.28(0.16) & 0.56(0.07)\\
  BZBJ1359-3746 & PMN J1359-3746 & 12.74(0.03) & 11.86(0.02) & 9.31(0.03) & 7.23(0.1) & 1.63(0.17) & 0.7(0.06)\\
  BZBJ1418-0233 & BZB J1418-0233 & 11.95(0.02) & 11.08(0.02) & 8.79(0.03) & 7.07(0.1) & 1.7(0.07) & 0.68(0.06)\\
  BZBJ1427+2348 & PKS 1424+240 & 10.22(0.02) & 9.38(0.02) & 7.08(0.02) & 5.21(0.03) & 1.78(0.02) & 0.59(0.05)\\
  BZBJ1439-1531 & PKS 1437-153 & 13.41(0.03) & 12.33(0.03) & 9.38(0.04) & 7.17(0.12) & 2.4(0.16) & 1.28(0.07)\\
  BZBJ1440+0610 & PMN J1440+0610 & 13.01(0.03) & 12.18(0.03) & 9.81(0.05) & 8.42(0.3) & 2.16(0.11) & 0.54(0.06)\\
  BZBJ1442+1200 & 1ES 1440+122 & 12.77(0.02) & 12.26(0.02) & 10.38(0.07) & 8.37() & 1.41(0.18) & -0.38(0.06)\\
  BZBJ1443-3908 & PKS 1440-389 & 11.7(0.02) & 10.97(0.02) & 8.79(0.03) & 6.89(0.08) & 1.77(0.06) & 0.24(0.06)\\
  BZBJ1501+2238 & MS 1458.8+2249 & 11.39(0.02) & 10.55(0.02) & 8.21(0.02) & 6.33(0.05) & 1.77(0.07) & 0.6(0.05)\\
  BZBJ1503-1541 & RBS 1457 & 13.74(0.03) & 13.12(0.03) & 10.93(0.13) & 9.08(0.54) & 1.8(0.15) & -0.08(0.07)\\
  BZBJ1506+0814 & PMN J1506+0814 & 12.9(0.03) & 12.15(0.02) & 9.98(0.05) & 7.87(0.16) & 1.96(0.16) & 0.34(0.06)\\
  BZBJ1516+1932 & PKS 1514+197 & 12.26(0.02) & 11.2(0.02) & 8.3(0.02) & 6.09(0.05) & 2.46(0.16) & 1.23(0.06)\\
  BZBJ1517-2422 & AP Librae & 10.0(0.02) & 9.11(0.02) & 6.46(0.02) & 4.34(0.02) & 2.05(0.04) & 0.72(0.06)\\
  BZBJ1522-2730 & PKS 1519-273 & 13.84(0.12) & 12.82(0.08) & 9.83(0.1) & 7.61(0.36) & 2.22(0.05) & 1.09(0.24)\\
  BZBJ1534+3715 & RGB J1534+372 & 13.45(0.03) & 12.81(0.03) & 10.84(0.09) & 8.7(0.37) & 2.15(0.16) & 0.0(0.07)\\
  BZBJ1540+8155 & 1ES 1544+820 & 12.9(0.02) & 12.2(0.02) & 10.15(0.04) & 8.26(0.17) & 1.48(0.16) & 0.16(0.06)\\
  BZBJ1546+0819 & 1RXS J154604.6+081912 & 13.58(0.03) & 12.76(0.03) & 10.59(0.08) & 8.97(0.42) & 1.57(0.21) & 0.51(0.07)\\
  BZBJ1548-2251 & PMN J1548-2251 & 12.49(0.02) & 11.76(0.02) & 9.64(0.04) & 7.78(0.15) & 1.93(0.13) & 0.26(0.04)\\
  BZBJ1552+0850 & TXS 1549+089 & 12.28(0.03) & 11.29(0.02) & 8.57(0.03) & 6.44(0.06) & 2.0(0.16) & 1.02(0.06)\\
  BZBJ1555+1111 & PG 1553+113 & 10.63(0.02) & 9.82(0.02) & 7.62(0.02) & 5.82(0.04) & 1.67(0.02) & 0.51(0.06)\\
  BZBJ1607+1551 & 4C +15.54 & 13.31(0.02) & 12.29(0.02) & 9.41(0.04) & 7.0(0.08) & 2.23(0.06) & 1.12(0.06)\\
  BZBJ1610-6649 & PMN J1610-6649 & 12.46(0.03) & 11.69(0.02) & 9.59(0.04) & 7.85(0.15) & 1.7(0.06) & 0.4(0.06)\\
  BZBJ1630+5221 & TXS 1629+524 & 13.63(0.03) & 12.81(0.02) & 10.66(0.06) & 8.8(0.26) & 2.03(0.1) & 0.52(0.06)\\
  BZBJ1642-0621 & TXS 1639-062 & 13.45(0.03) & 12.26(0.02) & 9.18(0.03) & 6.82(0.08) & 2.37(0.13) & 1.59(0.06)\\
  BZBJ1653+3945 & Mkn 501 & 9.86(0.02) & 9.4(0.02) & 7.34(0.02) & 5.4(0.03) & 1.74(0.03) & -0.52(0.05)\\
  BZBJ1719+1745 & PKS 1717+177 & 12.48(0.03) & 11.38(0.02) & 8.46(0.02) & 6.24(0.04) & 1.84(0.06) & 1.33(0.06)\\
  BZBJ1725+1152 & 1H 1720+117 & 11.77(0.02) & 10.95(0.02) & 8.76(0.03) & 6.97(0.08) & 1.93(0.06) & 0.53(0.06)\\
  BZBJ1725+5851 & 7C 1724+5854 & 12.59(0.02) & 11.68(0.02) & 9.2(0.03) & 7.33(0.08) & 2.26(0.17) & 0.79(0.06)\\
  BZBJ1728+5013 & I Zw 187 & 12.02(0.02) & 11.39(0.02) & 9.17(0.02) & 7.15(0.07) & 1.83(0.13) & -0.02(0.05)\\
  BZBJ1730+3714 & GB6 J1730+3714 & 13.14(0.02) & 12.41(0.02) & 10.17(0.04) & 8.18(0.18) & 2.09(0.14) & 0.27(0.06)\\
  BZBJ1739+4737 & S4 1738+47 & 13.26(0.02) & 12.18(0.02) & 9.25(0.03) & 6.94(0.06) & 2.09(0.15) & 1.26(0.06)\\
  BZBJ1742+5945 & RGB 1742+597 & 12.3(0.02) & 11.36(0.02) & 8.78(0.02) & 6.78(0.05) & 2.23(0.17) & 0.88(0.05)\\
  BZBJ1743+1935 & S3 1741+19 & 11.61(0.02) & 11.16(0.02) & 9.22(0.03) & 7.3(0.1) & 1.62(0.15) & -0.57(0.05)\\
  BZBJ1748+7005 & S4 1749+70 & 11.96(0.02) & 10.92(0.02) & 8.14(0.02) & 5.96(0.03) & 2.04(0.06) & 1.16(0.05)\\
  BZBJ1749+4321 & B3 1747+433 & 13.22(0.03) & 12.1(0.02) & 9.07(0.03) & 6.8(0.06) & 2.22(0.08) & 1.39(0.06)\\
  BZBJ1756+5522 & 1RXS J175615.5+552217 & 14.07(0.03) & 13.45(0.03) & 11.45(0.09) & 9.4(0.48) & 1.79(0.17) & -0.06(0.07)\\
  BZBJ1800+7828 & S5 1803+784 & 10.91(0.02) & 9.79(0.02) & 6.8(0.02) & 4.54(0.02) & 2.23(0.03) & 1.4(0.05)\\
  BZBJ1806+6949 & 3C 371 & 10.15(0.02) & 9.29(0.02) & 6.77(0.02) & 4.75(0.02) & 2.19(0.04) & 0.64(0.05)\\
  BZBJ1809+2910 & MG2 J180948+2910 & 12.33(0.02) & 11.37(0.02) & 8.77(0.03) & 6.74(0.06) & 2.04(0.11) & 0.95(0.06)\\
  BZBJ1813+3144 & B2 1811+31 & 12.4(0.02) & 11.61(0.02) & 9.4(0.04) & 7.72(0.15) & 2.11(0.07) & 0.45(0.06)\\
  BZBJ1813+0615 & TXS 1811+062 & 12.98(0.03) & 11.91(0.02) & 9.12(0.03) & 6.84(0.07) & 1.97(0.13) & 1.24(0.07)\\
  BZBJ1824+5651 & 4C +56.27 & 12.24(0.02) & 11.19(0.02) & 8.22(0.02) & 5.92(0.04) & 2.43(0.04) & 1.18(0.06)\\
  BZBJ1829+5402 & 1RXS J182925.7+540255 & 13.21(0.02) & 12.38(0.02) & 10.21(0.04) & 8.32(0.14) & 1.88(0.13) & 0.56(0.06)\\
  BZBJ1832-5659 & PMN J1832-5659 & 13.11(0.03) & 12.14(0.02) & 9.39(0.04) & 7.23(0.09) & 2.3(0.13) & 0.95(0.06)\\
  BZBJ1838+4802 & GB6 J1838+4802 & 12.87(0.03) & 12.11(0.02) & 10.0(0.04) & 8.04(0.12) & 1.72(0.1) & 0.34(0.06)\\
  BZBJ1849-4314 & PMN J1849-4314 & 11.93(0.03) & 10.97(0.02) & 8.4(0.02) & 6.34(0.05) & 2.02(0.09) & 0.93(0.06)\\
  BZBJ1917-1921 & 1H 1914-194 & 11.53(0.03) & 10.66(0.02) & 8.33(0.02) & 6.38(0.06) & 1.91(0.06) & 0.66(0.06)\\
  BZBJ1918-4111 & PMN J1918-4111 & 12.95(0.03) & 11.96(0.02) & 9.41(0.04) & 7.32(0.09) & 1.84(0.06) & 1.02(0.06)\\
  BZBJ1921-1607 & PMN J1921-1607 & 12.38(0.03) & 11.6(0.02) & 7.92(0.06) & 7.35(0.24) & 1.74(0.1) & 0.39(0.06)\\
  BZBJ1931+0937 & RX J1931.1+0937 & 11.79(0.04) & 11.12(0.03) & 9.26(0.04) & 7.57(0.12) & 2.36(0.07) & 0.09(0.08)\\
  BZBJ1936-4719 & PMN J1936-4719 & 13.52(0.03) & 12.88(0.03) & 10.78(0.09) & 8.73(0.37) & 1.64(0.16) & -0.0(0.07)\\
  BZBJ1945-3111 & PKS 1942-313 & 13.41(0.04) & 12.3(0.03) & 9.68(0.05) & 7.42(0.15) & 2.29(0.15) & 1.36(0.08)\\
  BZBJ2005+7752 & S5 2007+77 & 11.06(0.02) & 10.02(0.02) & 7.26(0.02) & 5.06(0.02) & 2.22(0.09) & 1.16(0.05)\\
  BZBJ2009-4849 & PKS 2005-489 & 10.23(0.02) & 9.49(0.02) & 7.23(0.02) & 5.36(0.03) & 1.78(0.05) & 0.29(0.05)\\
  BZBJ2009+7229 & 4C +72.28 & 13.16(0.02) & 12.06(0.02) & 9.21(0.03) & 6.88(0.05) & 2.3(0.08) & 1.35(0.06)\\
  BZBJ2015-0137 & PKS 2012-017 & 12.67(0.02) & 11.7(0.02) & 8.98(0.03) & 6.88(0.07) & 2.25(0.11) & 0.96(0.06)\\
  BZBJ2022+7611 & S5 2023+760 & 12.2(0.02) & 11.15(0.02) & 8.25(0.02) & 6.04(0.03) & 2.32(0.07) & 1.2(0.06)\\
  BZQJ0044-8422 & PKS 0044-84 & 13.73(0.03) & 12.65(0.03) & 9.81(0.03) & 7.44(0.09) & 2.53(0.11) & 1.28(0.06)\\
  BZQJ0102+4214 & CRATES J0102+4214 & 14.58(0.04) & 13.5(0.04) & 10.45(0.07) & 8.0(0.18) & 2.61(0.09) & 1.29(0.08)\\
  BZQJ0113+4948 & S4 0110+49 & 12.48(0.02) & 11.37(0.02) & 8.43(0.02) & 6.12(0.05) & 2.26(0.12) & 1.37(0.06)\\
  BZQJ0128+4439 & GB6 J0128+4439 & 14.82(0.04) & 13.87(0.04) & 11.72(0.21) & 8.68() & 2.25(0.13) & 0.9(0.1)\\
  BZQJ0136+4751 & OC 457 & 11.53(0.02) & 10.38(0.02) & 7.31(0.02) & 4.93(0.03) & 2.15(0.04) & 1.48(0.06)\\
  BZQJ0205+3212 & B2 0202+31 & 14.47(0.04) & 12.89(0.03) & 9.64(0.05) & 7.2(0.1) & 2.66(0.14) & 2.75(0.08)\\
  BZQJ0217+7349 & S5 0212+73 & 13.99(0.03) & 12.9(0.03) & 9.56(0.04) & 7.21(0.08) & 2.82(0.11) & 1.3(0.08)\\
  BZQJ0217+0144 & PKS 0215+015 & 11.41(0.02) & 10.26(0.02) & 7.28(0.02) & 4.96(0.03) & 2.15(0.03) & 1.47(0.05)\\
  BZQJ0230+4032 & B3 0227+403 & 13.78(0.03) & 12.58(0.03) & 9.66(0.04) & 7.43(0.12) & 2.63(0.06) & 1.62(0.07)\\
  BZQJ0237+2848 & 4C +28.07 & 12.51(0.02) & 11.37(0.02) & 8.43(0.03) & 6.09(0.05) & 2.16(0.06) & 1.44(0.06)\\
  BZQJ0245+2405 & B2 0242+23 & 15.48(0.05) & 14.45(0.07) & 11.09(0.14) & 8.63(0.34) & 2.54(0.08) & 1.14(0.15)\\
  BZQJ0250+1712 & NVSS J025037+171209 & 12.95(0.03) & 12.33(0.03) & 10.27(0.07) & 8.42(0.33) & 1.84(0.17) & -0.05(0.06)\\
  BZQJ0252-2219 & PKS 0250-225 & 14.28(0.03) & 13.04(0.03) & 9.81(0.05) & 7.38(0.12) & 2.19(0.05) & 1.76(0.07)\\
  BZQJ0257-1212 & PB 09399 & 14.26(0.03) & 13.08(0.03) & 10.17(0.05) & 8.09(0.17) & 2.39(0.14) & 1.57(0.07)\\
  BZQJ0303-7914 & PMN J0303-7914 & 14.28(0.03) & 13.06(0.03) & 10.02(0.04) & 7.53(0.09) & 2.2(0.13) & 1.7(0.07)\\
  BZQJ0309+1029 & PKS 0306+102 & 12.26(0.02) & 11.15(0.02) & 8.12(0.02) & 5.78(0.04) & 2.26(0.08) & 1.39(0.06)\\
  BZQJ0310+3814 & B3 0307+38 & 14.06(0.04) & 12.92(0.03) & 9.81(0.05) & 7.25(0.1) & 2.25(0.16) & 1.46(0.08)\\
  BZQJ0312+0133 & PKS 0310+013 & 13.51(0.03) & 12.41(0.03) & 9.61(0.04) & 7.05(0.08) & 2.26(0.08) & 1.34(0.06)\\
  BZQJ0315-1031 & PKS 0313-107 & 15.77(0.06) & 14.72(0.08) & 11.49(0.18) & 9.11() & 2.18(0.13) & 1.22(0.17)\\
  BZQJ0325+2224 & TXS 0322+222 & 13.68(0.03) & 12.42(0.03) & 9.16(0.04) & 6.75(0.09) & 2.41(0.12) & 1.82(0.07)\\
  BZQJ0336+3218 & NRAO 140 & 12.56(0.03) & 11.25(0.02) & 8.63(0.03) & 6.38(0.06) & 2.59(0.1) & 1.97(0.06)\\
  BZQJ0339-0146 & PKS 0336-01 & 12.7(0.03) & 11.54(0.02) & 8.58(0.03) & 6.16(0.05) & 2.48(0.07) & 1.51(0.06)\\
  BZQJ0348-2749 & PKS 0346-27 & 13.98(0.03) & 12.88(0.03) & 9.78(0.04) & 7.46(0.11) & 2.32(0.13) & 1.35(0.07)\\
  BZQJ0349-2102 & PKS 0347-211 & 15.59(0.05) & 14.42(0.06) & 11.29(0.13) & 8.57(0.3) & 2.23(0.09) & 1.55(0.13)\\
  BZQJ0402-3147 & PKS 0400-319 & 14.05(0.03) & 12.97(0.03) & 10.02(0.05) & 7.75(0.14) & 2.52(0.22) & 1.28(0.07)\\
  BZQJ0403-3605 & PKS 0402-362 & 11.49(0.02) & 10.27(0.02) & 6.99(0.02) & 4.52(0.02) & 2.3(0.04) & 1.69(0.06)\\
  BZQJ0405-1308 & PKS 0403-13 & 12.33(0.02) & 11.22(0.02) & 8.75(0.03) & 6.66(0.06) & 2.35(0.16) & 1.39(0.06)\\
  BZQJ0413-5332 & PMN J0413-5332 & 15.18(0.04) & 14.08(0.04) & 11.09(0.08) & 8.53(0.23) & 2.41(0.09) & 1.34(0.09)\\
  BZQJ0416-1851 & PKS 0414-189 & 14.86(0.04) & 13.56(0.04) & 10.35(0.06) & 7.99(0.17) & 2.2(0.09) & 1.92(0.09)\\
  BZQJ0422-0643 & PMN J0422-0643 & 12.56(0.02) & 11.58(0.02) & 8.78(0.03) & 6.54(0.06) & 2.39(0.12) & 0.99(0.06)\\
  BZQJ0423-0120 & PKS 0420-01 & 10.84(0.02) & 9.72(0.02) & 6.62(0.02) & 4.25(0.02) & 2.3(0.03) & 1.41(0.05)\\
  BZQJ0426+0518 & PKS 0423+051 & 14.79(0.04) & 13.52(0.04) & 10.59(0.08) & 8.22(0.24) & 2.66(0.12) & 1.85(0.09)\\
  BZQJ0438-1251 & PKS 0436-129 & 14.23(0.03) & 13.1(0.03) & 10.14(0.05) & 7.84(0.15) & 2.35(0.17) & 1.45(0.07)\\
  BZQJ0442-0017 & PKS 0440-00 & 12.9(0.02) & 11.84(0.02) & 8.94(0.03) & 6.6(0.05) & 2.44(0.03) & 1.22(0.06)\\
  BZQJ0448-2109 & PKS 0446-212 & 14.51(0.03) & 13.5(0.04) & 10.52(0.08) & 8.23(0.25) & 2.33(0.18) & 1.09(0.09)\\
  BZQJ0453-2807 & PKS 0451-28 & 13.49(0.03) & 12.4(0.02) & 9.15(0.03) & 6.59(0.06) & 2.66(0.05) & 1.32(0.06)\\
  BZQJ0455-4615 & PKS 0454-46 & 13.45(0.03) & 12.23(0.02) & 9.09(0.03) & 6.65(0.05) & 2.62(0.06) & 1.69(0.06)\\
  BZQJ0456-3136 & PMN J0456-3135 & 14.37(0.03) & 13.29(0.03) & 10.34(0.05) & 7.7(0.11) & 2.42(0.14) & 1.28(0.07)\\
  BZQJ0457-2324 & PKS 0454-234 & 12.27(0.02) & 11.14(0.02) & 8.15(0.02) & 5.81(0.03) & 2.03(0.02) & 1.41(0.06)\\
  BZQJ0501-0159 & S3 0458-02 & 13.19(0.03) & 12.01(0.02) & 8.89(0.03) & 6.44(0.06) & 2.52(0.1) & 1.59(0.06)\\
  BZQJ0502+0609 & PKS 0459+060 & 14.43(0.04) & 13.2(0.03) & 10.31(0.07) & 7.99(0.21) & 2.46(0.17) & 1.72(0.08)\\
  BZQJ0505-0419 & S3 0503-04 & 14.93(0.04) & 13.81(0.04) & 10.52(0.08) & 8.41(0.28) & 2.21(0.14) & 1.38(0.1)\\
  BZQJ0507-6104 & PMN J0507-6104 & 14.1(0.03) & 12.99(0.02) & 10.11(0.03) & 7.77(0.1) & 2.36(0.08) & 1.37(0.06)\\
  BZQJ0509+1011 & PKS 0506+101 & 14.26(0.03) & 13.23(0.03) & 10.45(0.08) & 7.89(0.19) & 2.33(0.09) & 1.14(0.08)\\
  BZQJ0510+1800 & PKS 0507+17 & 11.9(0.03) & 10.8(0.02) & 7.91(0.02) & 5.63(0.04) & 2.29(0.1) & 1.36(0.06)\\
  BZQJ0515-4556 & PKS 0514-459 & 12.73(0.02) & 11.76(0.02) & 8.93(0.02) & 6.56(0.04) & 2.47(0.18) & 0.96(0.05)\\
  BZQJ0526-4830 & PKS 0524-485 & 13.68(0.02) & 12.53(0.02) & 9.5(0.03) & 7.2(0.07) & 2.2(0.09) & 1.5(0.06)\\
  BZQJ0530+1331 & PKS 0528+134 & 14.7(0.04) & 13.65(0.04) & 10.5(0.09) & 7.74(0.16) & 2.22(0.09) & 1.18(0.1)\\
  BZQJ0529-0519 & PMN J0529-0519 & 14.3(0.03) & 13.42(0.04) & 10.19(0.06) & 7.83(0.17) & 2.3(0.39) & 0.71(0.09)\\
  BZQJ0532-3848 & PMN J0532-3848 & 15.34(0.04) & 14.11(0.04) & 11.04(0.09) & 8.62(0.27) & 2.61(0.15) & 1.73(0.1)\\
  BZQJ0532+0732 & OG 050 & 13.62(0.04) & 12.42(0.03) & 9.24(0.04) & 6.7(0.07) & 2.31(0.04) & 1.64(0.09)\\
  BZQJ0533+4822 & TXS 0529+483 & 12.7(0.03) & 11.45(0.02) & 8.26(0.02) & 5.8(0.04) & 2.31(0.05) & 1.78(0.06)\\
  BZQJ0539-2839 & PKS 0537-286 & 15.79(0.06) & 14.75(0.06) & 11.41(0.13) & 8.59(0.28) & 2.83(0.1) & 1.18(0.14)\\
  BZQJ0607-0834 & PKS 0605-08 & 11.94(0.02) & 11.36(0.02) & 9.06(0.03) & 6.81(0.06) & 2.36(0.08) & -0.19(0.06)\\
  BZQJ0610-6058 & PKS 0609-609 & 16.17(0.05) & 14.92(0.05) & 12.1(0.14) & 9.19(0.3) & 2.36(0.16) & 1.76(0.11)\\
  BZQJ0635-7516 & PKS 0637-75 & 11.57(0.02) & 10.49(0.02) & 7.65(0.02) & 5.36(0.03) & 2.65(0.06) & 1.28(0.05)\\
  BZQJ0701-4634 & PKS 0700-465 & 13.02(0.03) & 11.83(0.02) & 8.68(0.02) & 6.28(0.05) & 2.16(0.12) & 1.62(0.06)\\
  BZQJ0713+1935 & MG2 J071354+1934 & 11.87(0.02) & 10.84(0.02) & 8.1(0.02) & 5.89(0.04) & 2.01(0.06) & 1.13(0.05)\\
  BZQJ0725+1425 & 4C +14.23 & 13.42(0.03) & 12.31(0.03) & 9.43(0.04) & 6.98(0.08) & 2.04(0.04) & 1.37(0.06)\\
  BZQJ0726+2153 & TXS 0723+220 & 15.11(0.08) & 13.76(0.06) & 10.54(0.09) & 8.44(0.31) & 2.59(0.14) & 2.08(0.17)\\
  BZQJ0726-4728 & PMN J0726-4728 & 14.02(0.04) & 12.86(0.03) & 9.75(0.04) & 7.26(0.08) & 2.34(0.09) & 1.52(0.07)\\
  BZQJ0730-1141 & PKS 0727-11 & 11.91(0.02) & 10.76(0.02) & 7.68(0.02) & 5.28(0.04) & 2.11(0.02) & 1.48(0.05)\\
  BZQJ0733+5022 & TXS 0730+504 & 13.78(0.03) & 12.74(0.03) & 9.95(0.05) & 7.42(0.1) & 2.35(0.12) & 1.17(0.07)\\
  BZQJ0739+0137 & PKS 0736+01 & 11.04(0.02) & 10.02(0.02) & 7.24(0.02) & 5.03(0.03) & 2.23(0.08) & 1.11(0.05)\\
  BZQJ0746+2549 & B2 0743+25 & 16.11(0.08) & 14.9(0.09) & 11.39(0.17) & 8.48(0.28) & 2.85(0.11) & 1.66(0.2)\\
  BZQJ0749+4510 & B3 0745+453 & 11.98(0.03) & 10.95(0.02) & 8.07(0.02) & 5.51(0.03) & 2.24(0.16) & 1.16(0.06)\\
  BZQJ0750+1231 & OI 280 & 12.3(0.02) & 11.22(0.02) & 8.37(0.02) & 5.97(0.04) & 2.42(0.07) & 1.3(0.06)\\
  BZQJ0805+6144 & TXS 0800+618 & 15.59(0.05) & 14.4(0.06) & 11.02(0.1) & 8.4(0.26) & 2.74(0.07) & 1.62(0.13)\\
  BZQJ0808-0751 & PKS 0805-07 & 11.56(0.02) & 10.51(0.02) & 7.7(0.02) & 5.42(0.03) & 1.93(0.03) & 1.18(0.05)\\
  BZQJ0824+3916 & 4C +39.23 & 14.18(0.03) & 12.95(0.03) & 9.75(0.04) & 7.24(0.09) & 2.64(0.17) & 1.73(0.07)\\
  BZQJ0824+5552 & OJ 535 & 14.58(0.03) & 13.33(0.03) & 10.32(0.07) & 7.74(0.16) & 2.68(0.08) & 1.78(0.08)\\
  BZQJ0830+2410 & S3 0827+24 & 13.14(0.03) & 11.94(0.02) & 8.8(0.03) & 6.49(0.06) & 2.67(0.07) & 1.64(0.06)\\
  BZQJ0833+4224 & OJ 451 & 12.3(0.02) & 11.26(0.02) & 8.48(0.03) & 6.26(0.05) & 2.33(0.13) & 1.19(0.06)\\
  BZQJ0839+0104 & PKS 0837+012 & 14.92(0.04) & 13.68(0.04) & 10.66(0.09) & 8.18(0.25) & 2.21(0.11) & 1.75(0.1)\\
  BZQJ0841+7053 & 4C +71.07 & 13.73(0.03) & 12.59(0.03) & 9.16(0.03) & 6.68(0.06) & 2.95(0.07) & 1.47(0.06)\\
  BZQJ0903+4651 & S4 0859+47 & 14.36(0.03) & 13.15(0.03) & 9.98(0.05) & 7.58(0.12) & 2.27(0.18) & 1.67(0.08)\\
  BZQJ0912+4126 & B3 0908+416B & 15.18(0.04) & 14.0(0.05) & 10.94(0.09) & 8.12(0.17) & 2.3(0.17) & 1.59(0.1)\\
  BZQJ0916+3854 & S4 0913+39 & 14.57(0.03) & 13.43(0.04) & 10.5(0.08) & 8.39(0.29) & 2.53(0.16) & 1.47(0.09)\\
  BZQJ0920+4441 & S4 0917+44 & 13.59(0.03) & 12.36(0.03) & 9.09(0.03) & 6.58(0.07) & 2.11(0.03) & 1.73(0.07)\\
  BZQJ0921+6215 & OK 630 & 14.3(0.03) & 13.14(0.03) & 10.02(0.05) & 7.42(0.1) & 2.51(0.09) & 1.55(0.07)\\
  BZQJ0937+5008 & GB6 J0937+5008 & 13.19(0.03) & 12.02(0.02) & 8.91(0.03) & 6.4(0.06) & 2.5(0.15) & 1.55(0.06)\\
  BZQJ0957+5522 & 4C +55.17 & 12.91(0.03) & 11.85(0.02) & 8.94(0.03) & 6.63(0.06) & 1.83(0.03) & 1.22(0.06)\\
  BZQJ1044+8054 & S5 1039+81 & 12.44(0.02) & 11.33(0.02) & 8.42(0.02) & 6.12(0.04) & 2.54(0.15) & 1.39(0.06)\\
  BZQJ1056+7011 & S5 1053+70 & 14.86(0.03) & 13.62(0.03) & 10.22(0.05) & 7.7(0.12) & 2.64(0.1) & 1.74(0.08)\\
  BZQJ1058+8114 & S5 1053+81 & 13.37(0.03) & 12.2(0.02) & 9.04(0.03) & 6.64(0.06) & 2.58(0.09) & 1.57(0.06)\\
  BZQJ1258-2219 & PKS 1256-220 & 12.85(0.03) & 11.65(0.03) & 8.72(0.03) & 6.55(0.07) & 2.3(0.07) & 1.61(0.06)\\
  BZQJ1316-3338 & PKS 1313-333 & 12.77(0.03) & 11.65(0.02) & 8.68(0.03) & 6.34(0.06) & 2.31(0.06) & 1.41(0.06)\\
  BZQJ1332-1256 & PMN J1332-1256 & 15.51(0.06) & 14.29(0.07) & 11.36(0.22) & 8.1(0.27) & 2.38(0.04) & 1.69(0.16)\\
  BZQJ1337-1257 & PKS 1335-127 & 12.6(0.03) & 11.39(0.02) & 8.2(0.02) & 5.65(0.04) & 2.44(0.07) & 1.67(0.06)\\
  BZQJ1342-2051 & PKS B1339-206 & 14.72(0.04) & 13.26(0.04) & 10.05(0.06) & 7.58(0.14) & 2.63(0.13) & 2.4(0.09)\\
  BZQJ1344-1723 & PMN J1344-1723 & 13.33(0.03) & 12.19(0.03) & 9.15(0.04) & 6.9(0.09) & 1.95(0.06) & 1.47(0.06)\\
  BZQJ1347-3750 & PMN J1347-3750 & 14.47(0.04) & 13.32(0.04) & 10.33(0.06) & 7.94(0.17) & 2.32(0.12) & 1.46(0.09)\\
  BZQJ1351+0031 & PKS 1348+007 & 13.94(0.03) & 12.89(0.03) & 10.15(0.07) & 8.48(0.3) & 2.29(0.11) & 1.21(0.08)\\
  BZQJ1354-1041 & PKS 1352-104 & 13.42(0.03) & 12.4(0.03) & 9.3(0.04) & 6.96(0.09) & 2.57(0.08) & 1.09(0.07)\\
  BZQJ1357+7643 & S5 1357+76 & 14.95(0.03) & 13.76(0.03) & 10.79(0.06) & 7.99(0.13) & 2.3(0.1) & 1.63(0.07)\\
  BZQJ1408-0752 & PKS B1406-076 & 13.84(0.03) & 12.54(0.03) & 9.52(0.05) & 7.2(0.13) & 2.43(0.06) & 1.93(0.07)\\
  BZQJ1427-4206 & PKS B1424-418 & 11.26(0.02) & 10.17(0.02) & 7.22(0.02) & 4.94(0.03) & 1.96(0.03) & 1.32(0.05)\\
  BZQJ1436+2321 & PKS B1434+235 & 14.06(0.03) & 12.79(0.03) & 9.49(0.04) & 6.86(0.07) & 2.41(0.18) & 1.84(0.07)\\
  BZQJ1441-3303 & PKS 1438-328 & 14.17(0.03) & 13.04(0.04) & 10.01(0.06) & 7.62(0.15) & 2.76(0.17) & 1.44(0.08)\\
  BZQJ1443+2501 & PKS 1441+25 & 15.41(0.05) & 14.34(0.06) & 11.29(0.16) & 8.73(0.4) & 2.03(0.12) & 1.25(0.13)\\
  BZQJ1457-3539 & PKS 1454-354 & 13.0(0.03) & 11.92(0.02) & 9.07(0.03) & 6.89(0.07) & 2.11(0.03) & 1.28(0.06)\\
  BZQJ1504+1029 & PKS 1502+106 & 13.21(0.03) & 12.05(0.02) & 9.06(0.03) & 6.62(0.06) & 2.15(0.02) & 1.54(0.06)\\
  BZQJ1505+0326 & PKS 1502+036 & 13.94(0.03) & 12.95(0.03) & 9.68(0.04) & 7.03(0.09) & 2.51(0.07) & 1.0(0.07)\\
  BZQJ1506+3730 & B2 1504+37 & 13.9(0.03) & 12.54(0.03) & 9.56(0.04) & 7.45(0.1) & 2.57(0.1) & 2.09(0.07)\\
  BZQJ1509-4340 & PMN J1509-4340 & 14.27(0.05) & 13.14(0.04) & 10.5(0.07) & 7.74(0.14) & 2.65(0.14) & 1.44(0.11)\\
  BZQJ1510-0543 & PKS 1508-05 & 13.18(0.02) & 12.01(0.02) & 9.06(0.03) & 6.59(0.07) & 2.44(0.06) & 1.55(0.06)\\
  BZQJ1512-0905 & PKS 1510-08 & 11.27(0.02) & 10.17(0.02) & 7.37(0.02) & 5.06(0.03) & 2.29(0.01) & 1.32(0.06)\\
  BZQJ1520+4211 & B3 1518+423 & 13.15(0.02) & 12.21(0.02) & 9.48(0.03) & 7.3(0.09) & 2.5(0.31) & 0.88(0.06)\\
  BZQJ1521+4336 & B3 1520+437 & 15.62(0.05) & 14.5(0.06) & 10.98(0.1) & 8.74(0.31) & 2.99(0.16) & 1.4(0.13)\\
  BZQJ1539+2744 & MG2 J153938+2744 & 14.17(0.03) & 12.95(0.03) & 9.86(0.04) & 7.41(0.1) & 1.99(0.13) & 1.68(0.07)\\
  BZQJ1549+0237 & PKS 1546+027 & 12.21(0.02) & 11.13(0.02) & 8.47(0.03) & 6.22(0.05) & 2.46(0.07) & 1.27(0.06)\\
  BZQJ1550+0527 & 4C +05.64 & 13.38(0.03) & 12.2(0.02) & 9.1(0.03) & 6.61(0.07) & 2.32(0.11) & 1.6(0.06)\\
  BZQJ1608+1029 & 4C +10.45 & 12.56(0.03) & 11.46(0.02) & 8.6(0.03) & 6.23(0.04) & 2.33(0.1) & 1.34(0.06)\\
  BZQJ1610-3958 & PMN J1610-3958 & 11.98(0.03) & 10.87(0.02) & 7.95(0.02) & 5.58(0.03) & 2.61(0.1) & 1.36(0.07)\\
  BZQJ1613+3412 & OS 319 & 14.02(0.03) & 12.89(0.03) & 9.91(0.04) & 7.38(0.1) & 2.31(0.17) & 1.44(0.07)\\
  BZQJ1617-7717 & PKS 1610-77 & 13.23(0.03) & 12.16(0.03) & 9.07(0.03) & 6.62(0.06) & 2.5(0.05) & 1.27(0.06)\\
  BZQJ1635+3808 & 4C +38.41 & 13.09(0.03) & 11.86(0.02) & 8.51(0.02) & 5.96(0.04) & 2.25(0.03) & 1.71(0.06)\\
  BZQJ1637+4717 & 4C +47.44 & 12.63(0.02) & 11.45(0.02) & 8.35(0.02) & 5.98(0.04) & 2.41(0.06) & 1.59(0.06)\\
  BZQJ1640+3946 & NRAO 512 & 15.0(0.04) & 13.81(0.04) & 10.72(0.07) & 8.26(0.19) & 2.36(0.06) & 1.6(0.08)\\
  BZQJ1642+3948 & 3C 345 & 11.8(0.03) & 10.63(0.02) & 7.52(0.02) & 5.13(0.03) & 2.49(0.06) & 1.54(0.06)\\
  BZQJ1656+6012 & 87GB 165604.4+601702 & 14.26(0.03) & 13.26(0.03) & 10.62(0.05) & 8.18(0.14) & 2.36(0.21) & 1.03(0.06)\\
  BZQJ1703-6212 & CGRaBS J1703-6212 & 12.27(0.03) & 11.26(0.02) & 8.44(0.02) & 6.19(0.04) & 2.43(0.04) & 1.07(0.06)\\
  BZQJ1709+4318 & B3 1708+433 & 13.5(0.03) & 12.35(0.02) & 9.31(0.03) & 6.98(0.07) & 2.31(0.05) & 1.48(0.06)\\
  BZQJ1722+1013 & TXS 1720+102 & 12.42(0.02) & 11.33(0.02) & 8.33(0.02) & 5.98(0.04) & 2.23(0.06) & 1.31(0.06)\\
  BZQJ1724+4004 & S4 1722+40 & 14.16(0.03) & 13.09(0.03) & 10.0(0.04) & 7.66(0.1) & 2.34(0.06) & 1.25(0.06)\\
  BZQJ1727+4530 & S4 1726+45 & 13.67(0.02) & 12.51(0.02) & 9.37(0.03) & 6.82(0.06) & 2.58(0.06) & 1.52(0.06)\\
  BZQJ1728+1215 & PKS 1725+123 & 13.66(0.03) & 12.48(0.03) & 9.42(0.04) & 6.98(0.08) & 2.09(0.2) & 1.58(0.06)\\
  BZQJ1728+0427 & PKS 1725+044 & 12.3(0.02) & 11.31(0.02) & 8.45(0.03) & 5.87(0.04) & 2.53(0.08) & 1.02(0.06)\\
  BZQJ1730+0024 & PKS 1728+004 & 12.85(0.03) & 11.72(0.03) & 8.78(0.03) & 6.49(0.07) & 2.31(0.07) & 1.43(0.06)\\
  BZQJ1733-1304 & PKS 1730-13 & 12.14(0.03) & 11.06(0.02) & 8.1(0.02) & 5.66(0.04) & 2.24(0.09) & 1.3(0.06)\\
  BZQJ1734+3857 & B2 1732+38A & 13.49(0.03) & 12.31(0.02) & 9.15(0.03) & 6.79(0.06) & 2.24(0.04) & 1.58(0.06)\\
  BZQJ1739+4955 & S4 1738+49 & 12.61(0.03) & 11.5(0.02) & 8.64(0.02) & 6.37(0.04) & 2.2(0.09) & 1.36(0.06)\\
  BZQJ1740+5211 & 4C +51.37 & 12.33(0.02) & 11.16(0.02) & 8.16(0.02) & 5.89(0.03) & 2.5(0.04) & 1.54(0.05)\\
  BZQJ1745+2252 & TXS 1742+228 & 15.21(0.04) & 14.05(0.05) & 11.08(0.1) & 8.55(0.28) & 2.87(0.17) & 1.52(0.11)\\
  BZQJ1801+4404 & S4 1800+44 & 13.23(0.02) & 12.19(0.02) & 9.16(0.03) & 6.72(0.05) & 2.66(0.14) & 1.16(0.06)\\
  BZQJ1818+0903 & MG1 J181841+0903 & 13.08(0.03) & 11.98(0.03) & 9.11(0.03) & 6.82(0.08) & 2.32(0.08) & 1.32(0.07)\\
  BZQJ1848+3219 & B2 1846+32A & 13.36(0.03) & 12.19(0.02) & 9.13(0.03) & 6.69(0.06) & 2.38(0.09) & 1.54(0.06)\\
  BZQJ1852+4855 & S4 1851+48 & 13.08(0.03) & 12.0(0.02) & 9.13(0.02) & 6.89(0.06) & 2.28(0.04) & 1.28(0.06)\\
  BZQJ1903-6749 & PMN J1903-6749 & 13.46(0.03) & 12.38(0.02) & 9.49(0.03) & 7.08(0.08) & 2.49(0.1) & 1.3(0.06)\\
  BZQJ1911-2006 & PKS B1908-201 & 11.27(0.02) & 10.17(0.02) & 7.16(0.03) & () & 2.21(0.05) & 1.36(0.06)\\
  BZQJ1923-2104 & TXS 1920-211 & 11.09(0.02) & 10.08(0.02) & 7.33(0.02) & 5.02(0.03) & 2.1(0.04) & 1.08(0.05)\\
  BZQJ1924-2914 & PKS B1921-293 & 10.81(0.02) & 9.65(0.02) & 6.53(0.02) & 4.1(0.02) & 2.43(0.05) & 1.52(0.06)\\
  BZQJ1954-1123 & TXS 1951-115 & 13.68(0.04) & 12.5(0.03) & 9.45(0.05) & 7.08(0.12) & 2.25(0.05) & 1.57(0.08)\\
  BZQJ1957-3845 & PKS 1954-388 & 12.35(0.02) & 11.19(0.02) & 8.28(0.02) & 5.87(0.04) & 2.36(0.05) & 1.51(0.05)\\
  BZQJ1959-4246 & PMN J1959-4246 & 13.62(0.03) & 12.45(0.03) & 9.48(0.04) & 7.01(0.08) & 2.41(0.05) & 1.53(0.07)\\
  BZQJ2007-4434 & PKS 2004-447 & 13.37(0.03) & 12.28(0.03) & 9.42(0.04) & 7.01(0.09) & 2.47(0.12) & 1.3(0.06)\\
  BZQJ2023-1139 & PMN J2023-1140 & 15.38(0.05) & 14.63(0.08) & 11.24(0.17) & 7.96(0.21) & 2.07(0.11) & 0.32(0.16)\\
  BZQJ2025-0735 & PKS 2023-07 & 13.32(0.03) & 12.08(0.02) & 8.87(0.03) & 6.33(0.06) & 2.15(0.03) & 1.75(0.06)\\
  BZQJ2030-0622 & TXS 2027-065 & 14.15(0.03) & 12.93(0.03) & 9.8(0.06) & 8.03(0.26) & 2.73(0.1) & 1.7(0.08)\\
  BZQJ2056-4714 & PKS 2052-47 & 12.6(0.03) & 11.31(0.02) & 7.99(0.02) & 5.54(0.04) & 2.23(0.04) & 1.89(0.06)\\
  BZQJ2135-5006 & PMN J2135-5006 & 16.0(0.08) & 15.0(0.11) & 11.53(0.21) & 9.01() & 2.58(0.1) & 1.05(0.23)\\
  BZQJ2147-7536 & PKS 2142-75 & 11.01(0.02) & 9.78(0.02) & 6.61(0.02) & 4.23(0.02) & 2.52(0.04) & 1.72(0.05)\\
  BZQJ2202-8338 & PKS 2155-83 & 13.82(0.03) & 12.67(0.02) & 9.7(0.04) & 7.46(0.1) & 2.2(0.08) & 1.48(0.06)\\
\tableline  

\end{longtable}
\end{center}

{}

\end{document}